%% file: main.tex
\documentclass[a4paper,11pt]{article}
\usepackage{jcappub} 
\usepackage{lineno}

\usepackage{orcidlink}

\usepackage{array}

\usepackage{sectsty}
\allsectionsfont{\boldmath}

\usepackage{xspace}

\usepackage{tikz}
\usetikzlibrary{shapes.geometric, arrows}

\usepackage{float}

\usepackage{booktabs}
\usepackage{multirow}

\usepackage{graphicx}	
\usepackage{amsmath}	
\usepackage{mathtools}
\usepackage{physics}
\usepackage[dvipsnames]{xcolor}

\usepackage[capitalise,noabbrev,nameinlink]{cleveref}
\creflabelformat{equation}{#2#1#3} 
\usepackage{hyperref}

\newcommand{\btheta}{\boldsymbol{\theta}}

\newcommand{\pollux}{{\tt Pollux}\xspace}
\newcommand{\numba}{{\tt numba}\xspace}
\newcommand{\parsl}{{\tt Parsl}\xspace}
\newcommand{\healpix}{{\tt HEALPix}\xspace}
\newcommand{\healsparse}{{\tt HealSparse}\xspace}
\newcommand{\halofit}{{\tt HaloFit}\xspace}
\newcommand{\nside}{{\tt nside}\xspace}
\newcommand{\ccl}{{\tt CCL}\xspace}
\newcommand{\namaster}{{\tt NaMaster}\xspace}
\newcommand{\sacc}{{\tt SACC}\xspace}

\newcommand{\snapshotoutputs}{snapshot outputs\xspace}
\newcommand{\Snapshotoutputs}{Snapshot outputs\xspace}
\newcommand{\lightconeoutputs}{lightcone outputs\xspace}
\newcommand{\Lightconeoutputs}{Lightcone outputs\xspace}

\newcommand{\methodone}{Snapshot Slicing\xspace}
\newcommand{\methodtwo}{Snapshot Interpolation\xspace}
\newcommand{\methodthree}{Lightcone Slicing\xspace}

\newcommand{\zdownsampling}{\ensuremath{z_{\rm ds}}\xspace}
\newcommand{\nshells}{\ensuremath{N_{\rm shells}}\xspace}
\newcommand{\nsnapshots}{\ensuremath{N_{\rm snapshots}}\xspace}
\newcommand{\npart}{\ensuremath{N_{\rm part}}\xspace}

\newcommand{\ihMpc}{\ensuremath{\,h^{-1}\,\mathrm{Mpc}}\xspace}
\newcommand{\iMpc}{\ensuremath{\,\mathrm{Mpc}^{-1}}\xspace}
\newcommand{\hiMpc}{\ensuremath{\,h\,\mathrm{Mpc}^{-1}}\xspace}

\newcommand{\Gpc}{\ensuremath{\,\mathrm{Gpc}}\xspace}
\newcommand{\ihGpc}{\ensuremath{\,h^{-1}\,\mathrm{Gpc}}\xspace}

\usepackage{ulem}

\title{Optimization of Weak Lensing Lightcone Simulations for Higher-Order Statistics in the LSST era}
\input{DESC-PUB-00233_author_list}

\emailAdd{menafernandez@cppm.in2p3.fr}

\abstract{

We present a framework for generating lightcone simulations tailored to the analysis of Stage-IV cosmic shear data using Higher-Order Statistics (HOS). We revisit key design choices from previous simulation campaigns and re-optimize several internal parameters, benchmarking accuracy through changes in $\chi^2$ of cosmic shear statistics under survey conditions mimicking 10 years of observations from the Legacy Survey of Space and Time (LSST). We find that discretizing the lightcone uniformly in scale factor yields higher accuracy than commonly adopted schemes such as uniform spacing in redshift or comoving distance. While $\npart = 1024^3$ simulation particles (corresponding to a mass resolution of $m_{\rm part} = 2.08\times10^{10}M_\odot$) is sufficient to model two-point statistics up to $\ell = 5000$, we observed significant instabilities on our full suite of HOS as the number of mass shells used in the lightcone construction, $\nshells$, is varied. In contrast, simulations with $\npart = 2048^3$ particles ($m_{\rm part} = 2.60\times10^{9}M_\odot$) robustly reproduce all statistics considered. In this higher-resolution configuration, $\nshells$ can be reduced to $\sim50$ with only minor deviations—no larger than $0.1-0.3\sigma$ relative to our highest-resolution case ($\nshells\sim100$). This has been explicitly verified through a comparison between our fiducial lightcone production mode based on slicing particle snapshots and an exact lightcone mode where individual particle trajectories are solved for at runtime. We further show that the particle density per pixel can be downsampled by a significant amount for $z>1.5$, saving large computational resources with no impact on the resulting statistics. These results guide the design of upcoming simulation campaigns geared towards forward-modeling and emulation-based analyses of Stage-IV data. Alongside this work, we publicly release \pollux, the lightcone-construction framework developed for this project.

}

\begin{document}
\maketitle

\section{Introduction}

The next generation of astronomical surveys, such as the Vera C. Rubin Observatory Legacy Survey of Space and Time (LSST), promises to revolutionize our understanding of the Universe by delivering unprecedented depth, resolution and coverage of the sky \cite{lsst}. LSST is expected to observe billions of galaxies, enabling precision cosmological measurements through weak gravitational lensing and galaxy clustering \cite{mandelbaum2018weak}, among other probes. These probes are sensitive to both the geometry of the Universe and the growth of large-scale structure, making them essential tools for constraining the nature of dark energy, dark matter and potential modifications to General Relativity \cite{weinberg2013observational,amendola2018cosmology}.

Numerical simulations play a critical role in cosmological analyses. They provide realistic realizations of the large-scale structure of the Universe, including its non-linear and stochastic features \cite{heitmann2010coyote, hilbert2020accuracy}. Such simulations allow for the development and validation of analysis pipelines, the calibration of theoretical models, the quantification of systematic uncertainties, and underpin all simulation-based inference (SBI) techniques \cite{wu2013impact,mao2018descqa,jeffrey2025dark,von2025kids,novaes2025cosmology}. Unlike traditional methods, which rely on analytical approximations, SBI uses forward simulations to generate synthetic observables (implicit likelihood approaches) or to train surrogate models of these observables (emulators).

Emulator-based approaches have emerged as powerful tools for cosmological inference \cite{kwan2015cosmic,euclid2021euclid,angulo2021bacco} and use machine learning to approximate the outputs of complex simulations, enabling rapid predictions of observables given input cosmological parameters \cite{spurio2022cosmopower}. This strategy is particularly effective at handling the non-linearities present on small cosmological scales. By training on simulation outputs, emulators, such as those based on Gaussian processes or neural networks, allow for efficient reconstruction of posterior distributions with minimal information loss \cite{lawrence2017mira}. 

Recent progress has shown that emulator frameworks and physically-motivated models can effectively incorporate complex systematics, including intrinsic alignments and baryonic feedback, either by embedding these factors into the training data or by parametrizing them within halo-model–based approaches such as HMCode \cite{mead2021hmcode}. In addition, emulator-based methods are increasingly capable of synthesizing multiple cosmological probes, such as galaxy clustering and Cosmic Microwave Background (CMB) lensing, to produce tighter constraints on cosmological parameters. These advancements make emulator techniques highly promising for next-generation surveys such as LSST and \textit{Euclid} \cite{euclid}, where traditional approaches may not fully exploit the richness of the data.

While two-point correlation functions have been the cornerstone of cosmological analyses \cite{desy1_3x2pt,desy3_3x2pt,hamana2020cosmological,li2023hyper,asgari2021kids,wright2025kids}, Higher-Order Statistics (HOS) offer an approach to extracting information from the non-Gaussian features of the matter density field that two-point functions cannot access \cite{peebles1980large,bernardeau2002large}. Weak lensing and galaxy clustering at higher orders are particularly sensitive to the small-scale, non-linear regime of structure formation, which is rich in cosmological information but challenging to model accurately \cite{takada2003three}. This makes simulation-based approaches indispensable for robust theoretical predictions and error analysis. Previous cosmic shear simulation suites such as the cosmo-SLICS \cite{harnois2019cosmic}, CosmoGrid \cite{kacprzak2023cosmogridv1} and \textit{Gower Street} \cite{jeffrey2025dark} have enabled the development of lensing HOS and were successfully deployed in the analysis of Stage-III surveys \cite{heydenreich2022persistent,harnois2021cosmic,fluri2022full,marques2024cosmology,harnois2024kids,jeffrey2025dark}.

In this paper, we focus on the generation of weak lensing lightcone simulations tailored for higher-order statistical analyses in the LSST era. We explore different lightcone-slicing schemes---uniform in redshift, comoving distance and scale factor---and find that slicing uniformly in scale factor $a$ provides the most accurate results. We implement and compare three different algorithms for constructing the lightcones: \methodone, \methodtwo and \methodthree. The first selects particles within specific distance ranges from each simulation snapshot to construct the lightcone shells and serves as our baseline method; the second interpolates the lightcone positions of particles between adjacent snapshots and is tested as an alternative, though it yields less accurate results; and in the third, which we use to validate the baseline approach as it offers inherently better time resolution, the lightcone interpolation is performed dynamically at each time step during the production of the $N$-body simulation. 

Several existing large-scale lensing simulation efforts employ lightcone construction methods similar to those we test. The full-sky convergence maps developed for Hyper Suprime-Cam (HSC) analyses, and often referred to as ``T17'' \cite{takahashi2017full}, are generated by performing multiple-lens-plane ray tracing through a sequence of comoving shells extracted from discrete simulation snapshots, corresponding to our \methodone approach. In contrast, the \textit{Euclid} Flagship simulation constructs the lightcone \textit{on the fly} during the $N$-body run—interpolating particle positions at each time step to determine whether particles cross the observer's past lightcone—which matches our \methodthree approach \cite{castander2025euclid}.

Building on these approaches, we test the convergence of our lightcone construction as a function of the number of shells, \nshells, used to slice the simulation volume: increasing the number of shells improves time resolution and reduces discontinuities, but also increases computational and storage costs. We also study how simulation hyperparameters, particularly the mass resolution, affect our observables. The box size is fixed to $L_{\rm box}=600\ihMpc$ throughout the paper,\footnote{We plan to study the optimization of the box size in future work.} and we test two particle resolutions, $\npart=2048^3$ and $1024^3$, corresponding to mass resolutions of $m_{\rm part}=2.60\times10^9\,M_\odot$ and $2.08\times10^{10}\,M_\odot$, respectively. For both the simulations and the theoretical predictions, we use the LSST Science Requirements Document (SRD) source redshift distributions \cite{lsstsrd}.

Previous work investigated the effect of the number of shells and the mass resolution using flat-sky, pencil-beam simulations with modest particle resolution \cite{matilla2020optimizing}, providing early insights into the numerical requirements for lensing higher-order statistics. However, the simplified geometry and lower fidelity limit their applicability to Stage-IV surveys. Our work extends this line of research by employing high-resolution $N$-body simulations and explicitly testing the robustness of lightcone construction methods and simulation hyperparameters for LSST-level precision.

The outline of the paper is as follows: in \cref{sec:lensing}, we introduce the basics of weak lensing, where we include the expressions for the convergence and shear fields; in \cref{sec:methodology}, we present the $N$-body simulations used and introduce the methodology developed to build the lightcones (three different methods); in \cref{sec:results}, we present our results, where we test the convergence of our measurements from the lightcones with the number of shells and compare the outcomes of the different methodologies; and in \cref{sec:conclusions}, we highlight our conclusions.

\section{Weak lensing lightcones}\label{sec:lensing}

Weak gravitational lensing refers to the subtle distortion of the shapes of distant galaxies due to the gravitational influence of intervening mass distributions, such as galaxy clusters and large-scale structures along the line of sight. Unlike strong lensing, which produces highly visible phenomena such as multiple images and giant arcs, weak lensing results in small, coherent distortions that can only be detected statistically by averaging over many background galaxies \cite{Bartelmann2001, Kilbinger2015}.

The weak lensing formalism is based on the perturbed metric of general relativity, where the deflection of light is caused by the Newtonian potential $\Phi$ sourced by matter fluctuations. Under the Born approximation, which assumes that light rays travel along unperturbed paths, the lensing distortion is characterized by the Jacobian matrix:
\begin{align}\label{eq:jacobian}
    A_{ij} & =\pdv{\beta_i}{\theta_j}=\delta_{ij}-\pdv{\alpha_i}{\theta_j}\nonumber\\
    & =\delta_{ij}-\frac{2}{c^2}\int_0^\chi \dd{\chi'}\frac{f_K(\chi-\chi')f_K(\chi')}{f_K(\chi)}\pdv{\Phi\qty(f_K(\chi')\btheta,\chi')}{x_i}{x_j},
\end{align}
where $\btheta$ is the observed angular position, $\boldsymbol{\beta}$ is the unlensed source position, $\boldsymbol{\alpha}$ is the deflection angle, $f_K(\chi)$ is the comoving angular diameter distance, which depends on the curvature $K$ of the Universe, and $x_i$ is the $i$-th component of the transverse comoving position in the lens plane, $x_i=f_K(\chi)\theta_i$ \cite{Schneider2006}.

The deflection angle can be expressed as the gradient of the lensing potential $\psi$, defined as
\begin{equation}\label{eq:lensing_potential}
    \psi(\btheta,\chi)=\frac{2}{c^2}\int_0^\chi \dd{\chi'}\frac{f_K(\chi-\chi')}{f_K(\chi)f_K(\chi')}\Phi\qty(f_K(\chi')\btheta,\chi').
\end{equation}
Thus, the Jacobian matrix defined by \cref{eq:jacobian} takes the form
\begin{equation}
    A_{ij}=\delta_{ij}-\partial_i\partial_j\psi.
\end{equation}

The lensing distortion is commonly decomposed into two components: convergence $\kappa$, which describes isotropic magnification, and shear $\gamma$, which quantifies anisotropic stretching. The distortion matrix is given by
\begin{equation}
    \boldsymbol{A}=\mqty(1-\kappa-\gamma_1 & -\gamma_2\\-\gamma_2 & 1-\kappa+\gamma_1).
\end{equation}
The convergence and shear are related to the second derivatives of the potential as
\begin{align}
    \kappa &= \frac{1}{2}(\partial_1\partial_1+\partial_2\partial_2)\psi = \frac{1}{2}\boldsymbol{\nabla}^2_{\btheta}\psi, \label{eq:kappa_definition}\\
    \gamma_1 &= \frac{1}{2}(\partial_1\partial_1-\partial_2\partial_2)\psi, \\
    \gamma_2 &= \partial_1\partial_2\psi.
\end{align}
The convergence $\kappa$ represents a local magnification effect due to projected mass density, while the shear $\gamma$ quantifies the shape distortions induced by tidal gravitational fields \cite{Mellier1999, Munshi2008}. These quantities are fundamental for measuring the statistical properties of the large-scale structure with cosmic shear. In the next section, we describe how they are calculated in practice.

\subsection{Theoretical angular power spectrum}

\begin{figure}
    \centering
    \includegraphics[width=0.8\linewidth]{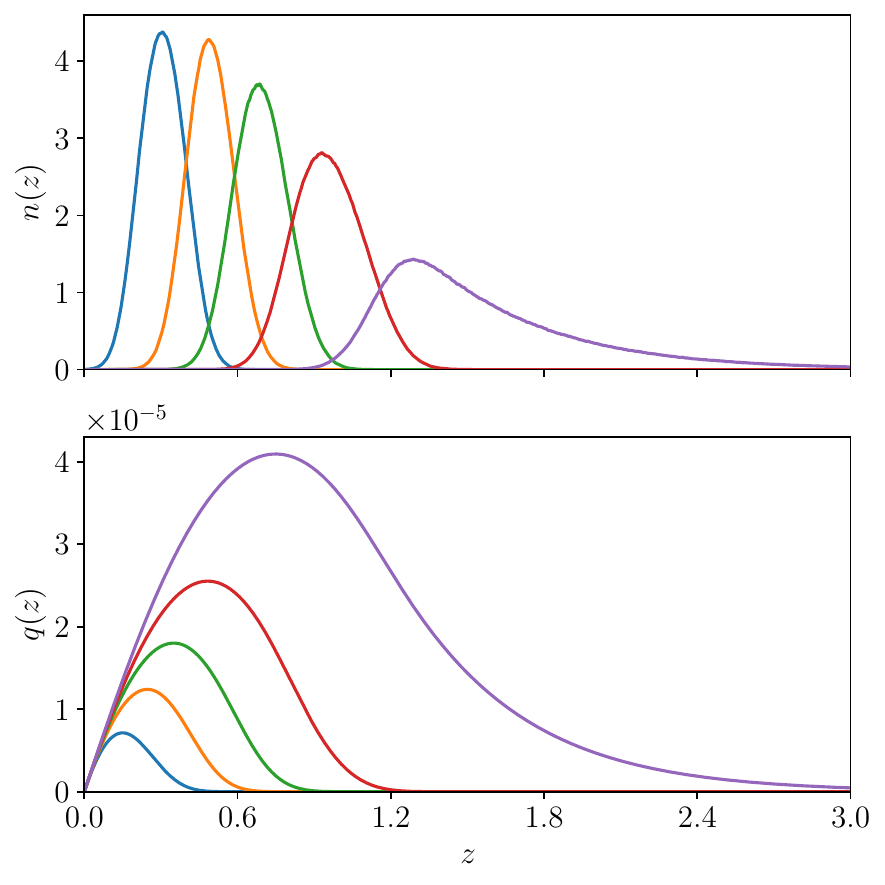}
    \caption{LSST Science Requirements Document (SRD) source redshift distributions for the different tomographic redshift bins (top), together with their corresponding lensing efficiencies (bottom).}
    \label{fig:nz_qz_SRD}
\end{figure}

Since the Newtonian potential is related to the matter overdensity field $\delta$ via the Poisson equation,
\begin{equation}
    \boldsymbol{\nabla}^2\Phi=\frac{3\Omega_{\rm m}H_0^2}{2a}\delta,
\end{equation}
using \cref{eq:lensing_potential,eq:kappa_definition} we find 
\begin{equation}\label{eq:kappa_thetachi}
    \kappa(\btheta,\chi) = \frac{3\Omega_{\rm m}H_0^2}{2c^2} \int_0^\chi \frac{\dd{\chi'}}{a(\chi')} \frac{f_K(\chi-\chi')}{f_K(\chi)}f_K(\chi')\delta(f_K(\chi')\btheta,\chi'),
\end{equation}
where $\Omega_{\rm m}$ is the matter density parameter, $H_0$ is the Hubble constant and $c$ is the speed of light in vacuum. The mean convergence $\kappa$ from a population of source galaxies is obtained by weighting the above expression with the galaxy probability distribution in comoving distance, $n(\chi)\dd{\chi}$,
\begin{equation}
    \kappa(\btheta)=\int_0^{\chi_{\rm lim}}\dd{\chi} \ n(\chi)\kappa(\btheta,\chi).
\end{equation}
The previous integral extends out to the limiting comoving distance of the galaxy sample, $\chi_{\rm lim}$. Using \cref{eq:kappa_thetachi}, we find
\begin{equation}
    \kappa(\btheta)=\int_0^{\chi_{\rm lim}}\dd{\chi}\ q(\chi)\delta(f_K(\chi)\btheta,\chi),
    \label{eq:kappa_theta}
\end{equation}
where $q(\chi)$ is the lensing efficiency,
\begin{equation}
    q(\chi)=\frac{3\Omega_{\rm m}H_0^2}{2c^2}\frac{f_K(\chi)}{a(\chi)}\int_\chi^{\chi_{\rm lim}}\dd{\chi'}n(\chi')\frac{f_K(\chi'-\chi)}{f_K(\chi')}.
\end{equation}
The LSST SRD source redshift distributions \cite{lsstsrd}, $n(z)$, related to $n(\chi)$ by $n(z)\dd{z} = n(\chi)\dd{\chi}$, are shown in \cref{fig:nz_qz_SRD} (top), together with their corresponding lensing efficiencies (bottom).

The shear field can be derived from the convergence. We begin by expanding $\kappa(\btheta,\chi)$, given in \cref{eq:kappa_thetachi}, in spherical harmonics,
\begin{equation}
    \kappa(\btheta,\chi) = \sum_{\ell m} \kappa_{\ell m}(\chi) Y_{\ell m}(\btheta),
\end{equation}
with coefficients
\begin{equation}
    \kappa_{\ell m}(\chi) = \int \dd{\Omega} \, \kappa(\btheta,\chi) Y_{\ell m}^*(\btheta).
\end{equation}
In harmonic space, these coefficients are related to those of the shear by
\begin{equation}\label{eq:gamma_ellm}
    \gamma_{\ell m}(\chi) = f_{\ell} \kappa_{\ell m}(\chi),
\end{equation}
where the filter function is defined as \cite{chang2018dark}
\begin{equation}
    f_{\ell} = -\sqrt{\frac{(\ell+2)(\ell-1)}{\ell (\ell+1)}}.
\end{equation}
We note that, therefore, in the absence of masking and noise, shear and convergence are related by an invertible linear transformation and therefore contain equivalent cosmological information. Since the shear is a spin-2 field, it is expanded in spin-weighted spherical harmonics as
\begin{equation}
    \gamma_1(\btheta,\chi) + \mathrm{i} \gamma_2(\btheta,\chi) = \sum_{\ell m} \gamma_{\ell m}(\chi) {}_{2}Y_{\ell m}(\btheta),
\end{equation}
where ${}_{2}Y_{\ell m}(\btheta)$ are the spin-2 spherical harmonics. The inverse transform then yields the real-space shear components $\gamma_1$ and $\gamma_2$.

Using the 3D Fourier transform of the matter field and assuming statistical isotropy and homogeneity, the matter power spectrum is defined by
\begin{equation}
    \expval{\delta(\boldsymbol{k}, \chi) \delta^*(\boldsymbol{k}', \chi')} 
    = (2\pi)^3 \delta_D(\boldsymbol{k} - \boldsymbol{k}') \delta_D(\chi - \chi') 
    P_{\rm m}(k, z(\chi)).
\end{equation}
The angular power spectrum of the convergence field between tomographic bins $a$ and $b$ is defined through
\begin{equation}
    \langle \kappa_{\ell m}^{(a)} \kappa_{\ell' m'}^{(b)*} \rangle 
    = \delta_{\ell \ell'} \delta_{m m'} \, C_\ell^{\kappa_a \kappa_b}.
\end{equation}
In the full expression for $C_\ell^{\kappa_a \kappa_b}$, the spherical Bessel functions of the first kind $j_\ell$ enter via the plane-wave expansion, leading to
\begin{equation}\label{eq:cell_bessel}
    C_\ell^{\kappa_a \kappa_b} 
    = \frac{2}{\pi} \int_0^{\chi_{\rm lim}} \dd{\chi}q^a(\chi) \int_0^{\chi_{\rm lim}} \dd{\chi'}q^b(\chi')
    \int \dd{k} \, k^2 P_{\rm m}(k, z) j_\ell(k\chi) j_\ell(k\chi').
\end{equation}
For large multipoles $\ell$, $j_\ell(k\chi)$ is strongly peaked around $k\chi \simeq \ell + \tfrac{1}{2}$ and can be approximated by
\begin{equation}
    j_\ell(k\chi) \simeq \sqrt{\frac{\pi}{2\ell+1}} \,
    \delta_D\!\left(k\chi - \left(\ell+\tfrac{1}{2}\right)\right).
\end{equation}
This is the essence of the Limber approximation \cite{limber1953analysis}: the Bessel kernel effectively enforces $k \simeq (\ell+1/2)/\chi$ inside the integral. Substituting this into \cref{eq:cell_bessel} collapses the $\chi'$ and $k$ integrals, yielding
\begin{equation}\label{eq:cell_limber}
    C_\ell^{\kappa_a \kappa_b} 
    \simeq \int_0^{\chi_{\rm lim}} \dd{\chi} \,
    \frac{q^a(\chi) q^b(\chi)}{\chi^2} \,
    P_{\rm m}\!\left(k=\frac{\ell+1/2}{\chi}, z(\chi)\right).
\end{equation}
This is the standard Limber form, accurate for multipoles $\ell\gtrsim20$ and slowly varying power spectra \cite{ loverde2008extended,kitching2017limits,kilbinger2017precision}.

\subsection{In practice: discretization of the lightcones}\label{sec:discretization}

In every approach to building lightcones, there are several key approximations that are either necessary or help simplify the calculations. Here, we focus on two of them: lightcone shell binning and assuming a static power spectrum within shells.\footnote{Additional approximations are discussed in \autoref{sec:approximations}.} To better understand the effect of these two approximations, we focus in this section on their impact on theoretical predictions, and defer the analysis using simulated particle densities to \cref{sec:methodology}.

\begin{figure*}
    \centering
    \includegraphics[width=\textwidth]{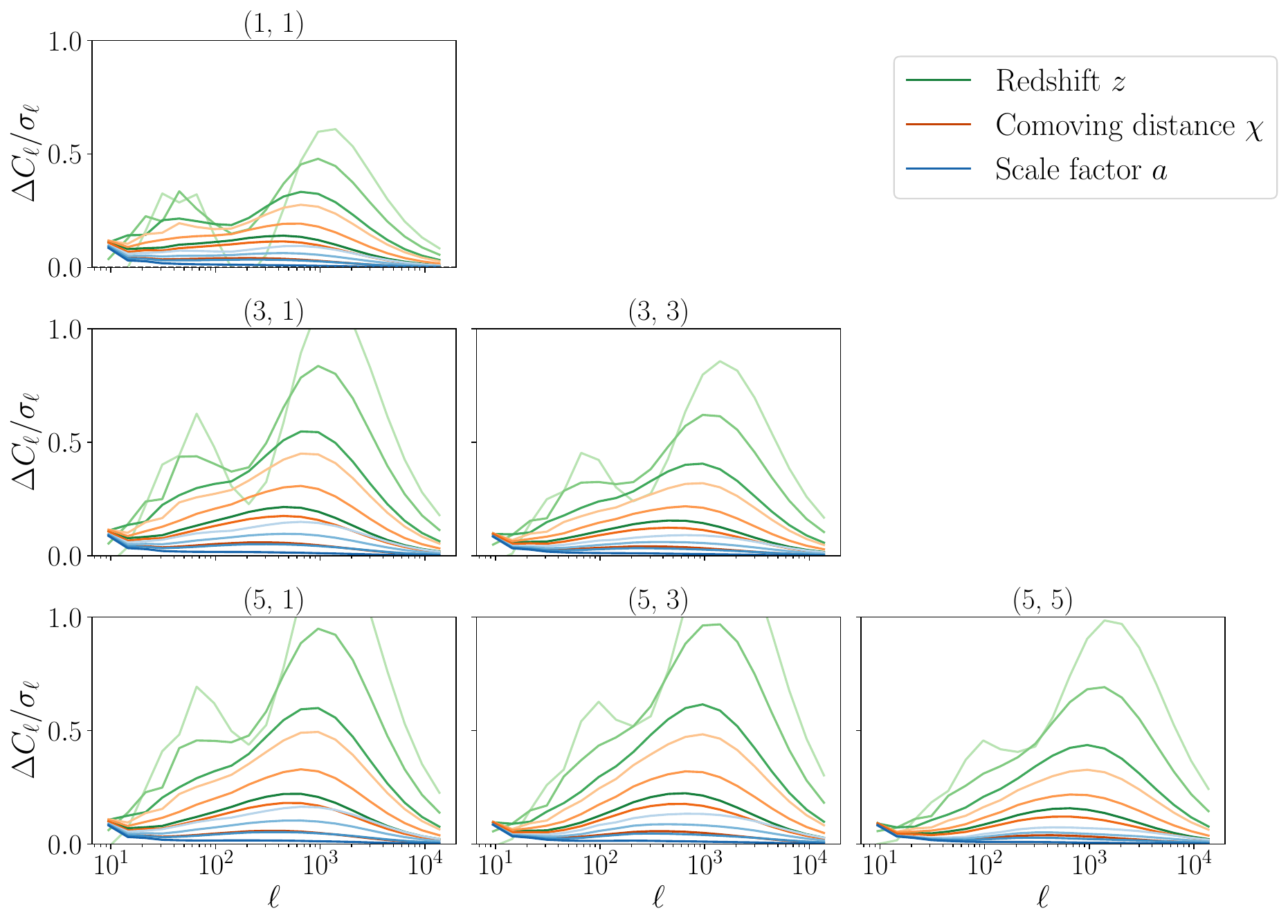}
    \caption{Relative difference in the $C_\ell$ of the convergence field for lightcone shells with edges linearly spaced in redshift $z$ (green), comoving distance $\chi$ (orange) and scale factor $a$ (blue) with respect to the full integral. The different color tones show the effect as a function of the number of shells \nshells used to discretize the lightcone in the radial direction (for $\nshells = 26$, 34, 51 and 101, from lighter to darker). Shells of constant width in scale factor yield better results. We use the SRD $n(z)$ to produce the $C_\ell$ and we normalize them by the diagonal of the Gaussian covariance matrix estimated for the LSST Y10 sample, described in \cref{sec:angular_power_spectrum}. We only show the results for the first, third and fifth tomographic bins.}
    \label{fig:binning_1}
\end{figure*}

\begin{figure*}
    \centering
    \includegraphics[width=\textwidth]{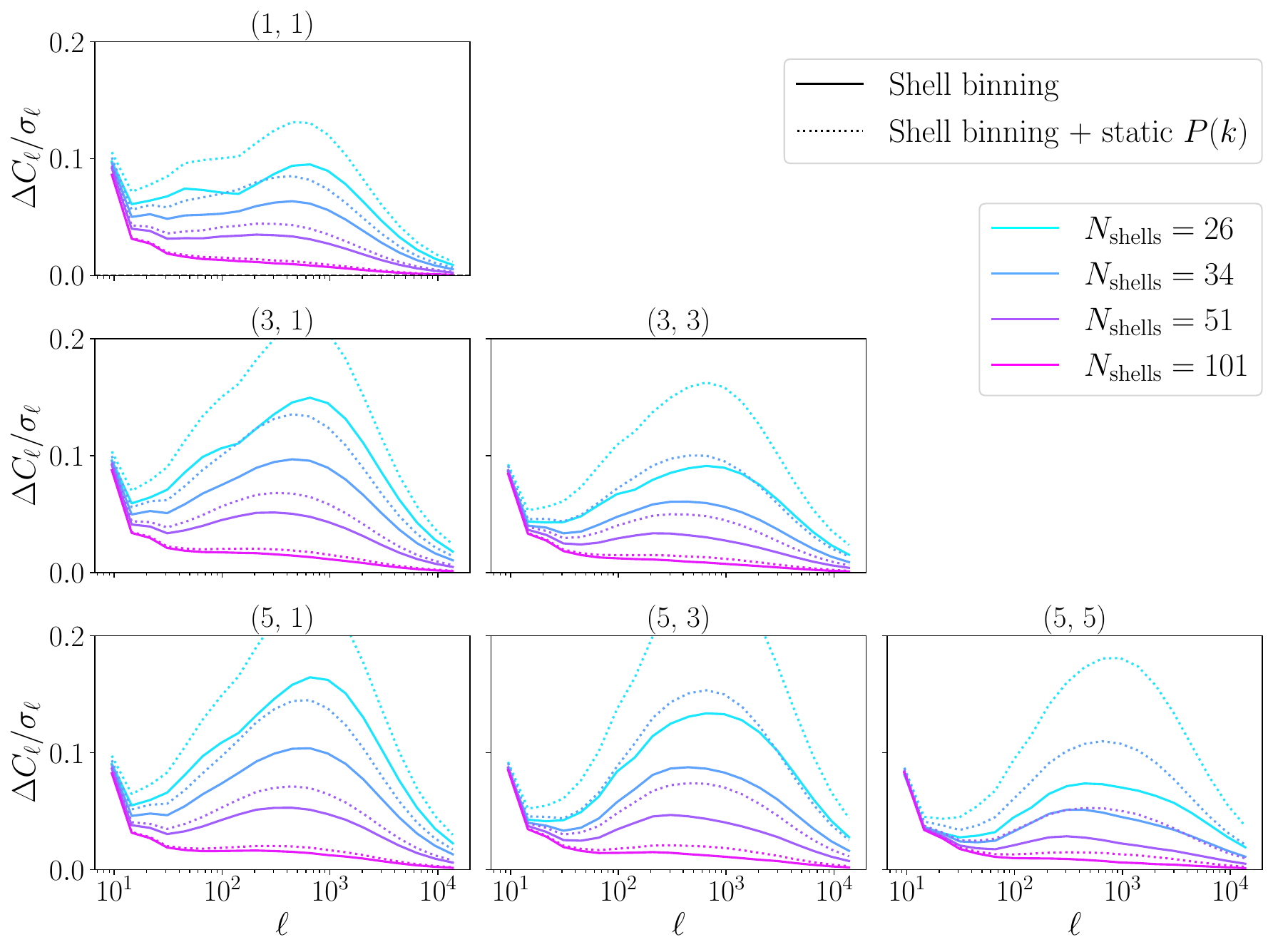}
    \caption{Effect of the lightcone shell binning (dashed) and non-evolving $P(k)$ (solid) in the $C_\ell$ of the convergence field. The plot shows the relative difference with respect to the full integration case. Different colors show different numbers of shells, uniformly sampled in scale factor. We only show the results for the first, third and fifth tomographic bins.}
    \label{fig:binning_2}
\end{figure*}

\textbf{Lightcone shell binning}. To produce lensing observables (convergence and shear) with ray-tracing algorithms, it is convenient to discretize the lightcone into a finite number of concentric shells around the observer and to compute the projected density of particles in each shell. For a given partition of the redshift range $[0,\, 4]$ into \nshells, \cref{eq:kappa_theta} may be rewritten, without any approximation, as
\begin{equation}
    \kappa(\btheta)=\sum_{i=1}^{\nshells} \int_{\chi_{i-1}}^{\chi_i} \dd{\chi}\ q(\chi)\delta(\chi\btheta,\chi),
\end{equation}
where the index $i$ denotes the $i$-th shell (we also assumed a spatially flat Universe, such that $f_K(\chi)=\chi$). Approximating the three-dimensional density field at distance $\chi$ by the projected density field within the corresponding $i$-th shell, ${\delta(\chi\btheta,\chi)\approx\delta_i(\btheta)}$ for ${\chi_{i-1} < \chi < \chi_i}$ (or, equivalently, taking $\delta$'s out of the integrals), this equation becomes
\begin{equation}\label{eq:kappa_stack}
    \kappa(\btheta)=\sum_{i=1}^{\nshells} w_i \delta_i(\btheta),
\end{equation}
where
\begin{equation}\label{eq:weights}
    w_i = \int_{\chi_{i-1}}^{\chi_i} \dd{\chi}\ q(\chi).
\end{equation}
Under these approximations, \cref{eq:cell_limber} becomes
\begin{equation}\label{eq:cell_limber_approx}
    C_\ell^{\kappa_a \kappa_b} = \sum_{i=1}^{\nshells} w^a_i w^b_i C_\ell^{\delta_i \delta_i},
\end{equation}
where the angular power spectrum of the overdensity field in shell $i$ is given by
\begin{equation}\label{eq:cell_density_limber}
    C_\ell^{\delta_i \delta_i}
    \simeq \int_{\chi_{i-1}}^{\chi_i} \frac{\dd{\chi}}{\chi^2} N(\chi)^2
    P_{\rm m}\qty(k=\frac{\ell+1/2}{\chi}, z(\chi)),
\end{equation}
where the spatial distribution of particles is proportional to the comoving volume, ${N(\chi)\propto\dv*{V}{\chi}}$. 

To evaluate differences between theoretical $C_\ell$ predictions, we first bin and convolve each one with the appropriate mode-coupling matrix to place them on the same footing as the simulated measurements performed on one octant of the sky.\footnote{All lightcones produced in this work cover one octant of the sky.} We then quantify their relative differences using the expected data uncertainties, characterized by the Gaussian covariance matrix for the binned $C_\ell$ estimated for the LSST Year 10 (Y10) sample, described in \cref{sec:angular_power_spectrum}.

\cref{fig:binning_1} shows the relative error, with respect to the full integral (\cref{eq:cell_limber}), for lightcone shells with edges linearly spaced in redshift $z$ (green), comoving distance $\chi$ (orange) or scale factor $a$ (blue), as a function of the number of shells (fixing the minimum and maximum redshifts to, respectively, 0 and 4). We consider four different values for the number of shells in which the lightcone volume is sliced: $\nshells=26$, 34, 51 and 101 (from light to dark), which matches the setup used for the simulations in \cref{sec:results}. We used the SRD source redshift distributions, shown in \cref{fig:nz_qz_SRD}, to produce these results. This figure shows that using shells of constant width in scale factor $a$ provides a more accurate approximation than adopting shells of fixed comoving distance or redshift width. Uniform binning in redshift performs poorly, yielding discrepancies exceeding $1\sigma$. Uniform splitting in comoving distance $\chi$ offers an improvement, and is commonly used in simulations (e.g., T17 \cite{takahashi2017full}, SLICS \cite{harnois2018cosmological} and CosmoDC2 \cite{korytov2019cosmodc2}) but remains suboptimal, with deviations of roughly $0.5\sigma$. Sampling uniformly in $a$ yields the best overall performance, with deviations below $0.2\sigma$.\footnote{Other simulation suites, such as \textit{Gower Street} \cite{jeffrey2025dark} and CosmoGrid \cite{kacprzak2023cosmogridv1}, adopt shells spaced uniformly in proper time. For clarity, we do not include this case in \autoref{fig:binning_1}; however, we found that proper-time binning performs intermediate between comoving-distance and scale-factor binning, therefore being suboptimal compared to scale factor.} In light of these results, we adopt this binning scheme for the remainder of the paper.

\textbf{Static power spectrum within shells}. In \cref{sec:methodology}, we consider various methods to project simulation particles onto the lightcone. In the simplest approach, we use the snapshot positions of particles, without correction, to populate each shell. This means that we treat the matter distribution as static across the shells, neglecting any time evolution. As a consequence, the power spectrum appears not to evolve within each shell. This approximation is useful for reducing computational complexity but may lead to inaccuracies in cases where shell-specific evolution is important. 

\cref{fig:binning_2} illustrates the combined impact of lightcone shell binning, using edges linearly spaced in scale factor and the assumption of a static power spectrum. This configuration closely mimics what would be obtained from a particle snapshot lightcone. For comparison, we also show the effect of lightcone shell binning alone (corresponding to the blue curves in \cref{fig:binning_1}). The static power spectrum is implemented by fixing the power spectrum to the snapshot redshift in \cref{eq:cell_density_limber}. As expected, including both effects leads to larger deviations from the full integral than accounting for shell binning alone, reaching up to $0.2\sigma$ for $\nshells = 26$, $0.08\sigma$ for $\nshells = 51$ and $0.03\sigma$ for $\nshells = 101$. These deviations decrease with increasing \nshells, consistent with the expectation in this purely theoretical setup that, in the limit of infinitely thin shells, we recover the full integral result.\footnote{We note that on very large angular scales ($\ell<15$), we found a difference of about 10\% of the error bars from the discretization, for all \nshells values. However, these predictions are computed under the Limber approximation, which is not valid on large scales for shells much thinner than the lensing window functions.} We note that, for simulations, using very thin shells could introduce discontinuity artifacts, motivating the numerical tests presented in the following sections.

\section{Simulation methodology}\label{sec:methodology}

In this section, we describe the methodology followed to build our lightcone simulations, from the $N$-body initial conditions to the production of galaxy mock catalogs, with a particular focus on the approximations assumed.

\subsection{HACC $N$-body simulation code}\label{sec:hacc}

The Hardware/Hybrid Accelerated Cosmology Code (HACC) is a high-performance simulation framework developed for large-scale cosmological studies \cite{habib2013hacc}. HACC is designed to run efficiently on a variety of supercomputing architectures, making it highly adaptable for current and next-generation high-performance computing systems. The code employs particle-based $N$-body methods to simulate the evolution of matter and large-scale structure in the Universe over cosmic time. Its versatility and efficiency have made it a cornerstone for cutting-edge cosmological simulations.

Among the major cosmological simulations run with HACC is the \textbf{OuterRim} simulation, which covers a volume of $(4.225\Gpc)^3$ and evolves over one trillion particles, achieving a mass resolution of about $2.60\times10^9\,M_\odot$ \cite{heitmann2019outer}. CosmoDC2, a large synthetic galaxy catalog covering 440 deg${}^2$ up to $z=3$ and developed to support precision dark energy science with the LSST, is based on the OuterRim run \cite{korytov2019cosmodc2}. Beyond OuterRim, other major HACC campaigns such as the \textbf{Last Journey} \cite{heitmann2021last}, \textbf{Farpoint} \cite{frontiere2022farpoint} and \textbf{New Worlds} \cite{heitmann2024new} simulations have extended to larger volumes, higher mass resolutions and varied cosmological models (including massive neutrinos and evolving dark energy), together providing rich datasets for precision cosmology.

One of HACC's key strengths lies in its hybrid architecture, which combines MPI for inter-node communication with multi-threading and GPU acceleration for intra-node computations \cite{heitmann2019hacc}. This hybrid approach enables HACC to scale effectively across thousands of compute cores, allowing it to handle simulations that resolve trillions of particles over cosmological volumes. 

The simulations are evolved from redshift 200 to redshift 0, and a total of 500 snapshots are stored, linearly spaced in scale factor. However, for our analysis, we focus only on snapshots corresponding to redshifts between 4 and 0, totaling 400 snapshots. Conveniently, HACC evolves simulations with constant steps in scale factor, which we showed is the optimal choice in \cref{fig:binning_1}. This means we can simply record snapshots every given number of steps to obtain a close-to-optimal redshift sampling.

HACC can store data in different ways:
\begin{itemize}
    \item \textbf{\Snapshotoutputs:} HACC stores selected snapshots while retaining either all simulation particles or a random selection at a given rate of downsampling. The frequency of stored redshift snapshots can be specified; in this work, we retain 101 snapshots between redshifts 4 and 0, corresponding to one-fourth of the total 400 snapshots. To post-process the outputs, we apply either the \methodone (method \ref{item:method1}) or the \methodtwo (method \ref{item:method2}) algorithm, which we describe in \cref{sec:algorithms}.
    
    \item \textbf{\Lightconeoutputs:} HACC interpolates between simulation time steps and retains only particles lying on the lightcone surface. When generating the particle lightcone, we first select the maximum redshift for which full particle outputs are required and, if relevant, we then specify the maximum redshift for downsampled outputs along with the corresponding downsampling rate. While this setup would produce 400 \lightconeoutputs between redshifts 0 and 4, for the tests presented in this paper we restrict the redshift range to $0 < z < 1$ to reduce computational costs, resulting in 252 \lightconeoutputs. These are subsequently post-processed using the \methodthree algorithm (method \ref{item:method3}), as described in \cref{sec:algorithms}.
\end{itemize}
All the simulations used in this paper have a box size of $L_{\rm box}=600\ihMpc$ and assume a flat-$\Lambda$CDM cosmology with the cosmological parameters listed in \cref{tab:cosmo_params}. We test two particle resolutions for the \snapshotoutputs ($\npart = 2048^3$ and $1024^3$) and a single one for the \lightconeoutputs ($\npart = 1024^3$).\footnote{We note that HACC adopts $N_{\rm mesh} = \npart$, where $N_{\rm mesh}$ denotes the number of grid cells per dimension used to compute gravitational forces (finer meshes provide higher force resolution). This is the standard choice in most cosmological particle-mesh codes. While this provides a good balance between computational cost and accuracy, it is possible that increasing the mesh resolution beyond the particle number could improve the convergence of certain statistics, such as the matter probability distribution function or higher-order moments, especially at low \npart, as discussed in \cite{uhlemann2020fisher}. Exploring the impact of a finer $N_{\rm mesh}$ could be an interesting direction for future work.}

\begin{table}
    \renewcommand{\arraystretch}{1.3} 
    \setlength{\tabcolsep}{4pt} 
    \centering
    \begin{tabular}{l c}
        \toprule\toprule
        Parameter & Value \\
        \hline
        $\Omega_{\rm c}$        & 0.22 \\
        $\Omega_{\rm b}$        & 0.0448 \\
        $h$                 & 0.71 \\
        $n_{\rm s}$             & 0.963 \\
        $\sigma_8$        & 0.80 \\
        $\sum m_\nu$ [eV]      & 0.0 \\
        \bottomrule\bottomrule
    \end{tabular}
    \caption{Cosmological parameters used for the simulations in this paper.}
    \label{tab:cosmo_params}
\end{table}

Given the size of the box and the number of particles, we can compute the Nyquist frequency as
\begin{equation}\label{eq:nyquist}
    k_{\rm Ny} = \pi\frac{\npart^{1/3}}{L_{\rm box}} = 
    \begin{cases}
        10.72\hiMpc, & \text{for } \npart=2048^3, \\
        5.36\hiMpc, & \text{for } \npart=1024^3,
    \end{cases}
\end{equation}
which sets the small-scale limit of our simulations. We can also compute the fundamental mode as
\begin{equation}
    k_{\rm min} = \frac{\pi}{L_{\rm box}} = 5.2\times10^{-3}\hiMpc,
\end{equation}
which corresponds to the largest-scale (smallest-wavenumber) mode that can be resolved in our simulations due to the finite size of the cubic box. It is also useful to define the corresponding minimum angular multipole, $\ell_{\rm min}$, which can be obtained from $k_{\rm min}$ via
\begin{equation}\label{eq:ellmin}
    \ell_{\rm min}(z_{\rm eff}) = \chi(z_{\rm eff}) k_{\rm min} = \frac{\pi \chi(z_{\rm eff})}{L_{\rm box}},
\end{equation}
where $z_{\rm eff}$ is the effective redshift. These quantities will be used to define the range of scales shown in the plots presented in \cref{sec:results}.

\subsection{Methods for building the lightcones}\label{sec:algorithms}

We test three different approaches to building the lightcones. Hereafter, we refer to them as \methodone (method \ref{item:method1}), \methodtwo (method \ref{item:method2}) and \methodthree (method \ref{item:method3}). All three methods discretize the lightcone into concentric shells around an observer and compute the projected density of particles in each of those shells. However, they differ in the way this discretization is performed and how particles are placed, given available snapshot information. Regardless of the method used, as mentioned earlier, all lightcones produced in this work cover an octant of the sky to reduce computational costs.

A common point between the three methods is that the original $N$-body boxes have to be replicated to tile the volume of the shells \cite{takahashi2017full}. When using HACC's \snapshotoutputs, we post-process them with methods \ref{item:method1} or \ref{item:method2}. We first replicate the box and apply an isometry to the specific grid of boxes, which have periodic boundary conditions. Doing so limits discontinuities in the lightcone density field. To avoid replication of structures along the line of sight, different random isometries, i.e., rotations and translations, are applied, one per group of adjacent shells of total depth not exceeding the box size. On the other hand, when using HACC's \lightconeoutputs, we post-process them with method \ref{item:method3}. In this case, the rotations and replications are applied by HACC itself with a more traditional approach compared to ours: the original box is replicated and then a random cube isometry is applied to each replica, consistent through time steps.

The three algorithms, which vary in their balance between computational complexity and accuracy, are described below:
\begin{enumerate}
    \item \textbf{\methodone.} This method post-processes HACC's \snapshotoutputs by selecting particles within specific distance ranges from simulation snapshots to construct lightcone shells. These shells are defined such that their boundaries (expressed in scale factor) are given by the mid scale factor between snapshots, effectively mapping each snapshot to a shell encompassing its corresponding redshift range. In this case, the positions of the particles at their snapshot redshifts are directly used. The freedom to apply random isometries (rotations and reflections) to the boxes allows us to produce several pseudo-independent realizations from the same $N$-body simulation. \label{item:method1}
     
    \item \textbf{\methodtwo.} This method post-processes HACC's \snapshotoutputs by interpolating the lightcone positions of particles between adjacent snapshots. The idea is to keep only those particles that cross the lightcone by linearly interpolating their radial comoving distance as a function of time (or scale factor), and finding if and where they cross the curve $\chi(a)$, obtained by integrating null geodesics. In this case, the shell boundaries correspond to the redshifts of the snapshots, as opposed to \methodone where the shells are ``centered'' around the snapshots. Similar to method \ref{item:method1}, it also relies on post-processing the \snapshotoutputs and allows us to produce several pseudo-independent realizations from the same simulation seed. \label{item:method2}
    
    \item \textbf{\methodthree.} This method post-processes HACC's \lightconeoutputs. In this approach, the lightcone interpolation is performed dynamically at each time step during the production of the $N$-body simulation, rather than as a post-processing of the raw \snapshotoutputs. This allows for a more precise determination of particles' positions and velocities as they cross the lightcone. However, even though this method has an inherently better time resolution compared to methods \ref{item:method1} and \ref{item:method2}, it has a major drawback: unlike in methods \ref{item:method1} and \ref{item:method2}, it does not allow us to apply the isometries to obtain several pseudo-independent realizations from the same $N$-body simulation (the rotations and replications are, in this case, applied ``in situ'' when running HACC, as discussed earlier). \label{item:method3}
\end{enumerate}

In \cref{tab:summary_hacc}, we summarize the settings used when building the lightcones with the three different approaches:
\begin{itemize}
    \item For method \ref{item:method1}, we test two different values for \npart ($2048^3$ and $1024^3$, yielding mass resolutions of $m_{\rm part} = 2.60\times10^9\,M_\odot$ and $m_{\rm part} = 2.08\times10^{10}\,M_\odot$, respectively) and four values for \nshells (26, 34, 51 and 101). Since it is the method that we test more extensively, we build the lightcones using five different seeds for the rotation of the boxes.
    
    \item For method \ref{item:method2}, we build the lightcones for $\npart = 2048^3$ and only for the $\nshells = 26$ and 101 cases, but using the same five seeds for the rotation of the boxes (to have a one-to-one comparison with method \ref{item:method1}). The reason behind using just $\nshells = 26$ and 101 is that we found method \ref{item:method2} to give comparable results to those of method \ref{item:method1}, but with slightly worse performance on small scales, as we show later.
    
    \item For method \ref{item:method3}, we only test the $\npart = 1024^3$ case, a single observer and only up to redshift 1. This is sufficient for our purposes, since method \ref{item:method3} is mainly intended to evaluate the accuracy of methods \ref{item:method1} and \ref{item:method2}, and the results obtained for $0 < z < 1$ are expected to remain valid at higher redshifts.
    
    \item For methods \ref{item:method1} and \ref{item:method2}, we use the five LSST SRD redshift distributions (shown in \cref{fig:nz_qz_SRD}) for the production of the convergence maps, since we have the snapshots covering the entire $0 < z < 4$ redshift range. For method \ref{item:method3}, we limit our study to the first three redshift bins (up to $z = 1$).
\end{itemize}

\begin{table*}
    \renewcommand{\arraystretch}{1.3} 
    \centering
    \begin{tabular}{ 
        >{\raggedright\arraybackslash}m{4.5cm} | 
        *{3}{>{\centering\arraybackslash}m{3.15cm}}
    }
        \toprule\toprule
        \textbf{Algorithm} & \textbf{\methodone} (\ref{item:method1}) & \textbf{\methodtwo} (\ref{item:method2}) & \textbf{\methodthree} (\ref{item:method3}) \\\hline
        HACC outputs & \Snapshotoutputs & \Snapshotoutputs & \Lightconeoutputs \\
        \npart & $2048^3$, $1024^3$ & $2048^3$ & $1024^3$ \\
        \nsnapshots available ($0<z<4$) & 101 & 101 & 400 \\
        \nshells tested & 26, 34, 51, 101 & 26, 101 & 26, 34, 51, 101, 201, 400 \\
        \nsnapshots/\nshells & 1 & 1 & $\geq 1$ \\
        Replications \& rotations & Applied during post-processing & Applied during post-processing & Applied ``in situ'' \\
        Number of pseudo-indep. realizations & 5 & 5 & 1 \\
        Redshift distributions (redshift range) & SRD ($0<z<4$) & SRD ($0<z<4$) & SRD ($0<z<1$) \\
        \bottomrule\bottomrule
    \end{tabular}
    \caption{Summary of the settings used when post-processing the HACC simulations with our different algorithms. Both methods \ref{item:method1} and \ref{item:method2} use HACC's \snapshotoutputs, whereas method \ref{item:method3} post-processes HACC's \lightconeoutputs. Regardless of the method used, all lightcones produced for this work cover an octant of the sky.}
    \label{tab:summary_hacc}
\end{table*}

\subsection{Computational cost mitigation strategy}\label{sec:approximations}

Besides the lightcone shell binning and the fact that the power spectrum is non-evolving within shells for method \ref{item:method1} (discussed in \cref{sec:discretization}),\footnote{This is not true for method \ref{item:method3}, which has sub-shell evolution. Method \ref{item:method2} aims at recovering evolution by interpolating between fixed-time snapshots.} there are other key approximations that allow us to effectively simulate large volumes while keeping the computational costs manageable. Reducing computational cost is crucial for our purposes, as we aim to generate a large number of independent realizations to robustly estimate the variance and covariance of higher-order weak-lensing statistics. Moreover, we need to produce simulations at different cosmologies to enable parameter inference with a simulation-based approach, which further amplifies the computational demands. Achieving this requires minimizing the cost per realization without compromising the physical accuracy of our simulations.

\begin{itemize}

    \item \textbf{Downsampling particles.} Keeping all particles makes building the lightcones an extremely computationally expensive task, especially at high redshift ($z\gtrsim2$), where the lensing and clustering kernels decrease rapidly (see the bottom plot of \cref{fig:nz_qz_SRD}). Because of this, we implemented a downsampling algorithm. The way we downsample is the following: we fix the average pixel density for all shells above redshift \zdownsampling to the value it had at that redshift (see the left panel of \cref{fig:downsampling}). In the right panel of \cref{fig:downsampling}, we show the effect of this downsampling methodology for different values of \zdownsampling: 1.5 (our baseline choice\footnote{The choice of $\zdownsampling = 1.5$ as our baseline value is justified in \cref{app:downsampling}, where we argue that it has little impact on the angular power spectrum for the scales we consider.}), 0.8 and 0.5. For $\zdownsampling = 1.5$, roughly 50\% of the simulation particles are downsampled at $z \sim 2$, increasing to about 75\% by $z = 4$. 
    \begin{figure*}
        \centering
        \begin{minipage}[t]{0.49\textwidth}
            \centering
            \includegraphics[width=\textwidth]{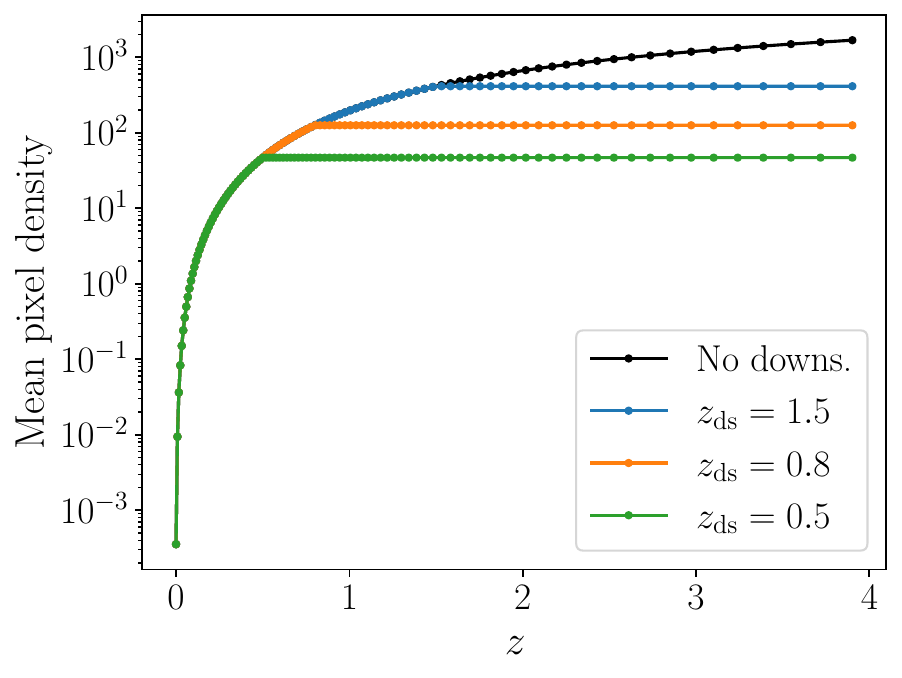}
        \end{minipage}
        \hfill
        \begin{minipage}[t]{0.49\textwidth}
            \centering
            \includegraphics[width=\textwidth]{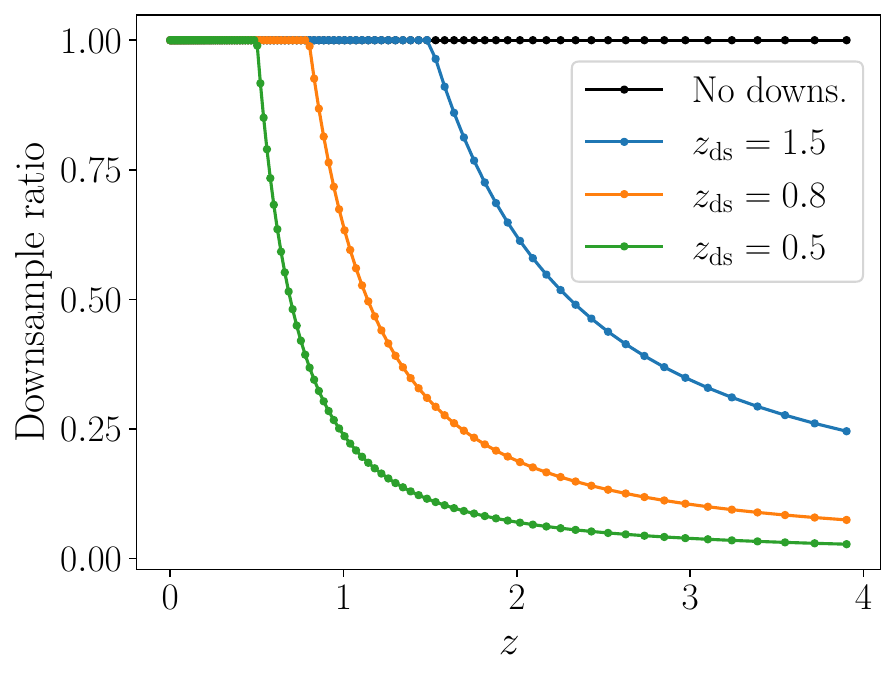}
        \end{minipage}
        \caption{\textbf{Left panel:} mean pixel density as a function of redshift for one of our HACC simulations. The black curve represents all particles without any downsampling, whereas blue, orange and green show the result of downsampling from redshifts 1.5, 0.8 and 0.5, respectively. \textbf{Right panel:} Downsample ratio as a function of redshift for one of our HACC simulations. We show the ratio of the number of particles with and without downsampling for the same three values of \zdownsampling as in the left panel, which is a direct measure of the resulting speedup.}
        \label{fig:downsampling}
    \end{figure*}
    
    \item \textbf{Born approximation.} We compute the weak lensing convergence and shear fields using the Born approximation, which assumes that light rays follow unperturbed paths through the matter distribution. The Born approximation has been shown to provide a generally accurate description of weak-lensing two-point and higher-order statistics when compared to full ray tracing \cite{petkova2013glamer,takahashi2017full,petri2017validity,ferlito2024ray}. After discretizing the lightcone, we produce two kinds of convergence maps: the first ones are tomographic $\kappa$ maps, which mimic the convergence estimated from source galaxies that follow the SRD $n(z)$; the second ones are shell $\kappa$ maps, which correspond to a sequence of source planes defined at the upper redshift edge of each shell. In order to compute both, we stack the projected overdensity field computed in each shell following \cref{eq:kappa_stack} with weights given by \cref{eq:weights}. Alternatively, these weights are usually computed under the multiple-lens-plane approximation (see, e.g., \cite{teyssier2009full,sgier2019fast,vecchi2025impact,ferlito2023millenniumtng}) whereby, in each shell, all particles are projected onto a source plane at an effective redshift \cite{harnois2015simulations}.\footnote{Given the mass of the particles $m_{\rm part}$ and area $A_{\rm pix}$ of each pixel (in steradian), the effective surface mass density of shell $i$ is given by $$\hat{\Sigma}_i(\btheta)=\frac{m_{\rm part}}{A_{\rm pix}d_A^2(z_i^l)}\hat{N}_i(\btheta),$$ where $d_A$ is the angular diameter distance and $\hat{N}_i$ is the particle count map of that shell. Here, it is assumed that the whole mass of the shell is located at the mean redshift of the particles, $z_i^l$. For a series of lens planes, the convergence is given by $$\kappa(\btheta, z_s) = \sum_l \Sigma_l(\btheta) / \Sigma_{\rm crit}(z_l,z_s),$$ where $\Sigma_{\rm crit}$ is the critical surface mass density. This expression can be integrated over the redshift distribution of the sources, as needed. The convergence then reads as $\kappa(\btheta) = \sum_i w_i \hat{N}_i(\btheta)$ with corresponding weights finally given by $$w_i= \frac{m_{\rm part}}{A_{\rm pix} d_A^2(z_i^l)} \int \dd{z_s}\frac{n(z_s)}{\Sigma_{\rm crit}(z_i^l,z_s)}.$$} We found that the two computations provide extremely close results, which is expected for thin shells \cite{teyssier2009full}.

    \item \textbf{Paired-seed averaging strategy.} To reduce the impact of sample variance in our statistical measurements, we adopt a paired-seed averaging strategy similar to that used in \cite{harnois2019cosmic, harnois2023mglens}. Specifically, we identify two random realizations of the initial conditions whose individual power spectra fluctuate in opposite directions relative to the linear-theory predictions. By averaging the two corresponding spectra, the leading-order sample variance approximately cancels out, yielding a result that more closely tracks the theoretical expectation. This technique, related to the ``paired-and-fixed'' approach \cite{angulo2016cosmological}, has been shown to significantly suppress cosmic variance in $N$-body simulations, enabling accurate recovery of clustering and lensing statistics with fewer realizations. The effectiveness of the paired-seed averaging procedure in reducing sample variance is illustrated in \cref{fig:pk_seeds_average}. The figure shows the power spectra of the two individual seeds, selected from 200 realizations as the pair whose average best matches the theoretical prediction, together with their average, all normalized to the theory. The averaged spectrum (blue line) is significantly less noisy than the individual seeds (orange and green) and stays within 5\% of the theoretical prediction over the range $2\times10^{-2}\,h\,{\rm Mpc}^{-1} < k < 3\,h\,{\rm Mpc}^{-1}$.
    
    \begin{figure}
        \centering
        \includegraphics[width=\linewidth]{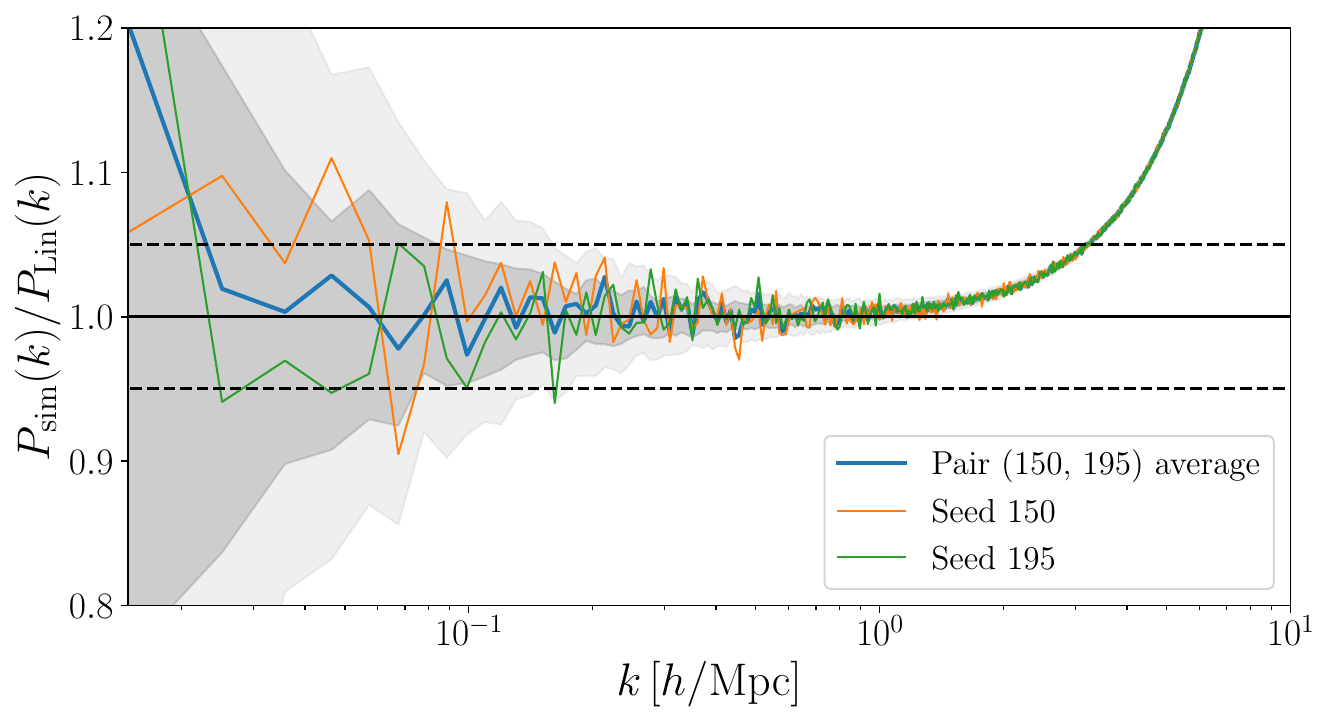}
        \caption{Power spectrum ratio relative to linear theory. The blue line shows the average of the two random seeds that best match the theoretical prediction, while the orange and green lines correspond to the individual seeds. The shaded regions indicate the scatter across the ensemble of 200 realizations: the darker band corresponds to the $1\sigma$ interval, and the lighter band to the $2\sigma$ one.}
        \label{fig:pk_seeds_average}
    \end{figure}

    \item \textbf{Finite box size.} Since our simulations use a box of finite comoving size, $L_{\rm box}$, they inherently lack density fluctuations with wavelengths larger than this scale. This limitation prevents accurate modeling of structure formation on the largest scales and alters the covariance properties \cite{sirko2005initial}. To cover the full lightcone volume, especially at high redshifts, we replicate and rotate the simulation boxes, as detailed previously. While these operations help avoid visual repetition and populate the volume, they do not reintroduce the missing long-wavelength modes. To visualize the scales impacted by these missing modes, most plots in \cref{sec:results} include gray-shaded regions indicating scales larger than the size of the box, corresponding to multipoles $\ell \leq \ell_{\rm min}$, defined by \cref{eq:ellmin}.
    
\end{itemize}

\subsection{Implementation in \texttt{Pollux}}
\label{sec:pollux}

\pollux\footnote{\url{https://github.com/LSSTDESC/pollux}.} is a lightcone–construction framework developed for this project to efficiently convert $N$-body simulation outputs into weak-lensing and large-scale-structure observables. It provides a unified, modular interface for post-processing particle data produced either as discrete snapshots or as on-the-fly lightcone outputs, and includes dedicated readers for both HACC and {\sc cubep$^3$m}\footnote{\url{https://github.com/jharno/cubep3m}.} \cite{harnois2013high,harnois2018cosmological} simulations. \pollux implements all three lightcone-building methods examined in this work---\methodone (\ref{item:method1}), \methodtwo (\ref{item:method2}) and \methodthree (\ref{item:method3})---and provides a common pipeline for the ensuing tasks of shell construction, particle projection and map generation. \pollux relies on the LSST Core Cosmology Library (\ccl) \cite{chisari2019core} for all theoretical quantities, including cosmological distances. A brief summary of the internal module structure of \pollux is provided in \cref{app:pollux}.

A key feature of \pollux is its fully Python implementation combined with just-in-time compilation (via \numba\footnote{\url{https://github.com/numba/numba}.}) to accelerate the most computationally expensive operations, such as particle selection, coordinate transformations and spherical projection. These tasks are automatically parallelized and executed as independent modules, enabling \pollux to scale efficiently over many compute nodes. Workflow orchestration is handled through \parsl,\footnote{\url{https://github.com/Parsl/parsl}.} which allows each small, localized task (e.g., processing a single subvolume or shell) to be distributed across available resources with minimal overhead.

Given $N$-body snapshot or lightcone outputs, \pollux first defines the set of concentric shells that discretize the past lightcone.\footnote{The shell centers correspond to the snapshot redshifts, and their edges are defined as the average redshift between two adjacent snapshots.} When snapshot outputs are used (methods \ref{item:method1} and \ref{item:method2}), \pollux replicates simulation boxes and applies random isometries to populate each shell volume, enabling multiple pseudo-independent realizations from a single simulation, as described in \cref{sec:algorithms}. For lightcone outputs (method \ref{item:method3}), \pollux instead stacks the pre-selected particles delivered by HACC, preserving the time resolution inherent to the on-the-fly construction.

Within each shell, \pollux projects particles onto HEALPix pixelizations---typically at high resolution (\nside = 8192) but configurable---producing sparse, memory-efficient particle count maps using \healsparse.\footnote{\url{https://github.com/LSSTDESC/healsparse}.} These projected maps are then converted into overdensity fields via
\begin{equation}
    \delta = \frac{\npart - \bar{N}_{\rm part}}{\bar{N}_{\rm part}},
\end{equation}
where the expected mean count $\bar{N}_{\rm part}$ is computed analytically from the simulation and cosmology, thereby avoiding additional noise introduced by empirical estimates. Weak-lensing quantities are then obtained by stacking the shells under the Born approximation, using the lensing-efficiency weights of \cref{eq:weights} to generate both per-shell and tomographic convergence and shear maps, as described in \cref{sec:approximations}. \pollux can also produce mock galaxy catalogs designed to follow the given tomographic redshift distributions $n(z)$ by sampling the density field with an optional linear bias\footnote{\pollux supports several angular sampling schemes: (i)~uniform; (ii)~clustered sampling based on the density field using multinomial draws (fixed total number of galaxies); and (iii)~clustered sampling using Poisson draws (fixed galaxy density).} and interpolating the non-tomographic $\kappa$ and $\gamma$ values at each galaxy’s position and redshift. At last, \pollux can post-process halo catalogs identified in the snapshots to construct corresponding halo lightcone catalogs, which will be exploited in future work.

In summary, \pollux provides a flexible and computationally efficient environment for constructing weak-lensing lightcones from large $N$-body simulations. Its modular design, high-performance implementation and ability to handle multiple construction methods make it well suited for generating the large ensembles of simulations required for higher-order weak-lensing statistics and simulation-based inference.

\subsection{Statistics used for method comparison}\label{sec:stats}

We have presented three different methodologies for constructing the lightcones. In \cref{sec:results}, we assess their convergence and mutual consistency using the angular power spectrum ($C_\ell$) and higher-order moments of $\kappa$. In this section, we describe the procedures used to measure these statistics from our $\kappa$ maps.

\subsubsection{Angular power spectrum}\label{sec:angular_power_spectrum}

The angular power spectrum, $C_\ell$, is measured from the simulations using the publicly available code \namaster\footnote{\url{https://github.com/LSSTDESC/NaMaster}.} with 20 logarithmically spaced $\ell$ bins in the range ${8 < \ell < 2\times\nside}$ and stored as \sacc\footnote{\url{https://github.com/LSSTDESC/sacc}.} FITS files. Theoretical predictions are processed through \namaster as well, ensuring that they are consistently convolved with the mode-coupling matrix and binned identically to the measurements.

We compute Gaussian covariance matrices for the $C_\ell$, assuming LSST Y10 error bars. The theoretical predictions are obtained with \ccl, employing the \halofit prescription to model the non-linear matter power spectrum \cite{smith2003halofit, takahashi2012halofit, mead2015halofit}, although we also explore other alternatives for the power spectrum in \cref{sec:results_theory}. The covariance matrix is calculated using the Knox formula \cite{knox1995determination}.
\begin{equation}\label{eq:cov}
    {\rm Cov}_{qq'}^{(ij),(mn)} = 
    \delta_{qq'}\frac{1}{(2\ell_q+1)\,f_{\rm sky}\,n_q}
    \left[D_q^{im} D_q^{jn} + D_q^{in} D_q^{jm}\right],
\end{equation}
$\ell_q$ is the mean multipole of the $q$-th bandpower, $n_q$ is the number of multipoles contributing to it, $\delta_{qq'}$ is the Kronecker delta function and $f_{\rm sky}$ is the sky fraction used in the analysis. As discussed in \cref{sec:algorithms}, we set $f_{\rm sky} = 1/8$, corresponding to one octant of the sky. The quantity $D_q^{ij}$ is defined as
\begin{equation}\label{eq:D_q}
    D_q^{ij} =
    \begin{cases}
        C_q^{ij} + N_q^i, & \text{if } i=j, \\
        C_q^{ij}, & \text{if } i \neq j,
    \end{cases}
\end{equation}
where $N_q^i$ denotes the shot noise, modeled as white (scale-independent) noise:
\begin{equation}
    N_q^i = \frac{\sigma_\epsilon^2}{\bar{n}^i}.
\end{equation}
Here, $\sigma_\epsilon = 0.3$ is the root-mean-square intrinsic ellipticity per component and $\bar{n}^i = 5$ gal$/$arcmin${}^2$ is the mean galaxy number density per steradian for the tomographic bin $i$ (for LSST Y10, a redshift-combined value of $\bar{n} \simeq 25$ gal$/$arcmin${}^2$ is expected for the sources, see \cite{chang2013effective}).

The covariance matrix of \cref{eq:cov} is also used to evaluate the $\chi^2$ between angular power spectra measured from different lightcone realizations with varying numbers of shells, which will be useful for us to assess their convergence in \cref{sec:results}. The $\chi^2$ is given by
\begin{equation}\label{eq:chi2}
    \chi^2 = \sum_{ij}\sum_{mn}\sum_{qq'}
    \left(C_q^{(ij),A} - C_q^{(ij),B}\right)
    \left[{\rm Cov}^{-1}\right]_{qq'}^{(ij),(mn)}
    \left(C_{q'}^{(mn),A} - C_{q'}^{(mn),B}\right),
\end{equation}
where $i$, $j$, $m$ and $n$ denote tomographic bins, and $A$ and $B$ label the lightcone configurations with different values of $\nshells$. All the $\chi^2$ computed in this work use \cref{eq:ellmin} for the minimum multipole included in their calculation.

\subsubsection{Higher-order moments of $\kappa$}

From our simulations, we measure the second, third and fourth central moments of the $\kappa$ field, and check the convergence as a function of \nshells for each of them. The second moment of the convergence as a function of $\vartheta$ contains information similar to standard shear two-point statistics. However, the third and fourth moments contain additional non-Gaussian information. The second and third moments were used by the Dark Energy Survey to constrain the amplitude of matter fluctuations $S_8$ using its first three years of data in \cite{gatti2022dark}.

We follow the next steps to compute them:
\begin{enumerate}
    \item The convergence maps are smoothed by a top-hat filter of smoothing length $\vartheta$, $\kappa(\btheta)\to\kappa_\vartheta(\btheta)$. We do this for the different redshift bins (i.e., we actually have $\kappa_\vartheta^i(\boldsymbol{\theta})$, where $i$ indicates the tomographic redshift bin), using a set of eight logarithmically spaced smoothing scales in the range $\vartheta \in [2.5\,\mathrm{arcmin},250\,\mathrm{arcmin}]$.
    \item The second central moment (or covariance) is computed as
    \begin{equation}
        \langle\kappa^2_\vartheta\rangle^{ij} = 
            \langle
            (\kappa^i_\vartheta(\btheta) - \langle\kappa^i_\vartheta(\btheta)\rangle) \cdot (\kappa^j_\vartheta(\btheta) - \langle\kappa^j_\vartheta(\btheta)\rangle)
            \rangle;
    \end{equation}
    the third one (related to the skewness), as
    \begin{equation}
            \langle\kappa^3_\vartheta\rangle^{ijk} = 
            \langle
            (\kappa^i_\vartheta(\btheta) - \langle\kappa^i_\vartheta(\btheta)\rangle)
            \cdot (\kappa^j_\vartheta(\btheta) - \langle\kappa^j_\vartheta(\btheta)\rangle)
            \cdot (\kappa^k_\vartheta(\btheta) - \langle\kappa^k_\vartheta(\btheta)\rangle)
            \rangle;
    \end{equation}
    and the fourth one (related to the kurtosis), as
    \begin{equation}\label{eq:fourth_moment}
        \begin{aligned}
            \langle\kappa^4_\vartheta\rangle^{ijkl} = 
            \langle &
            (\kappa^i_\vartheta(\btheta) - \langle\kappa^i_\vartheta(\btheta)\rangle)
            \cdot (\kappa^j_\vartheta(\btheta) - \langle\kappa^j_\vartheta(\btheta)\rangle) \\
            &\cdot (\kappa^k_\vartheta(\btheta) - \langle\kappa^k_\vartheta(\btheta)\rangle)
            \cdot (\kappa^l_\vartheta(\btheta) - \langle\kappa^l_\vartheta(\btheta)\rangle)
            \rangle.
        \end{aligned}
    \end{equation}
    The angle brackets $\langle \cdot \rangle$ denote an average over sky positions $\boldsymbol{\theta}$ (i.e., a spatial average across the field).
\end{enumerate}

\section{Results}\label{sec:results}

In this section, we explore different configurations for building our lightcone simulations, as detailed in \cref{sec:algorithms}, and check their performance with respect to the angular power spectrum and higher-order moments of $\kappa$, which we measure following \cref{sec:stats}. Our baseline approach is the \methodone algorithm (method \ref{item:method1}) with $\npart = 2048^3$, which we validate in more detail in \cref{sec:results_method1}. We also test its consistency with methods \ref{item:method2} and \ref{item:method3} in \cref{sec:results_method2,sec:results_method3}, respectively. Finally, we also assess the level of agreement with the expected theoretical predictions in \cref{sec:results_theory}.

\begin{figure*}
    \centering
    \begin{minipage}[t]{0.47\textwidth}
        \centering
        \includegraphics[width=\textwidth]{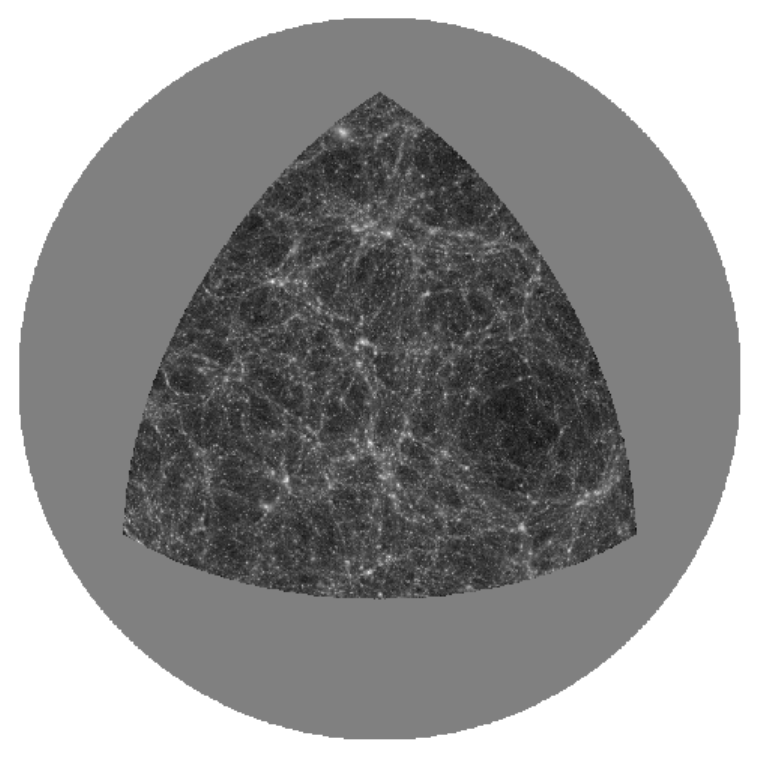}
    \end{minipage}
    \hfill
    \begin{minipage}[t]{0.47\textwidth}
        \centering
        \includegraphics[width=\textwidth]{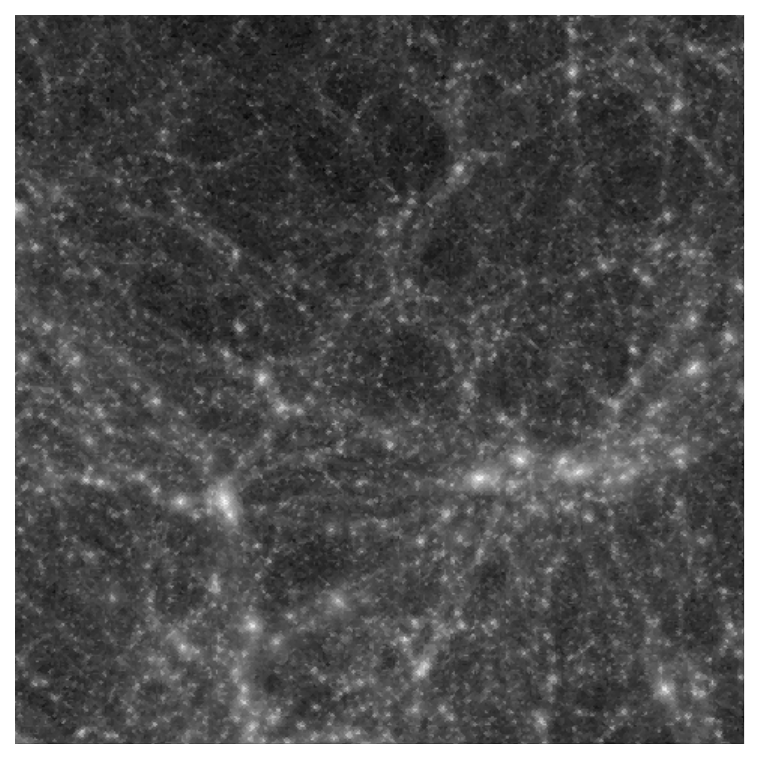}
    \end{minipage}
    \caption{Example of the (log-scaled) $\delta$ map of a lightcone shell for one of our simulations. For this particular shell, the redshift slice is given by $z\in(0.016, 0.050)$, corresponding to the $z=0.033$ snapshot. \textbf{Left panel:} orthographic projection of the entire octant, centered at (45 deg longitude, 37.5 deg latitude). \textbf{Right panel:} cartesian projection of a $30\times30$ deg${}^2$ region of the octant near the equator.}
    \label{fig:map_density_shell}
\end{figure*}

As an example, in \cref{fig:map_density_shell}, we show the $\delta$ map of a lightcone shell for one of our simulations. The left panel illustrates the entire octant while the right one shows a zoom on a $30\times30$ deg${}^2$ region.

\subsection{Results for \methodone}\label{sec:results_method1}

In this section, we take a detailed look at the \methodone algorithm (method \ref{item:method1}), where we study its convergence with the number of shells and the impact of the number of particles on its results.

\subsubsection{Convergence with \nshells}\label{sec:convergence_method1}

Our goal here is to study the convergence of our simulations, generated with the \methodone algorithm (method \ref{item:method1}), as we increase the number of shells in which these are binned (or, equivalently, as we reduce their width). Specifically, we use $\nshells = 26$, 34, 51 and 101 shells, corresponding to one-fourth, one-third, one-half and the full set of available snapshots, respectively. We use simulations with $\npart = 2048^3$ and average all the results over the two simulation seeds available and a total of five random seeds for the observers per simulation seed, i.e., over a total of ten data vectors.

\paragraph{Angular power spectrum}

\begin{figure*}
    \centering
    \includegraphics[width=\textwidth]{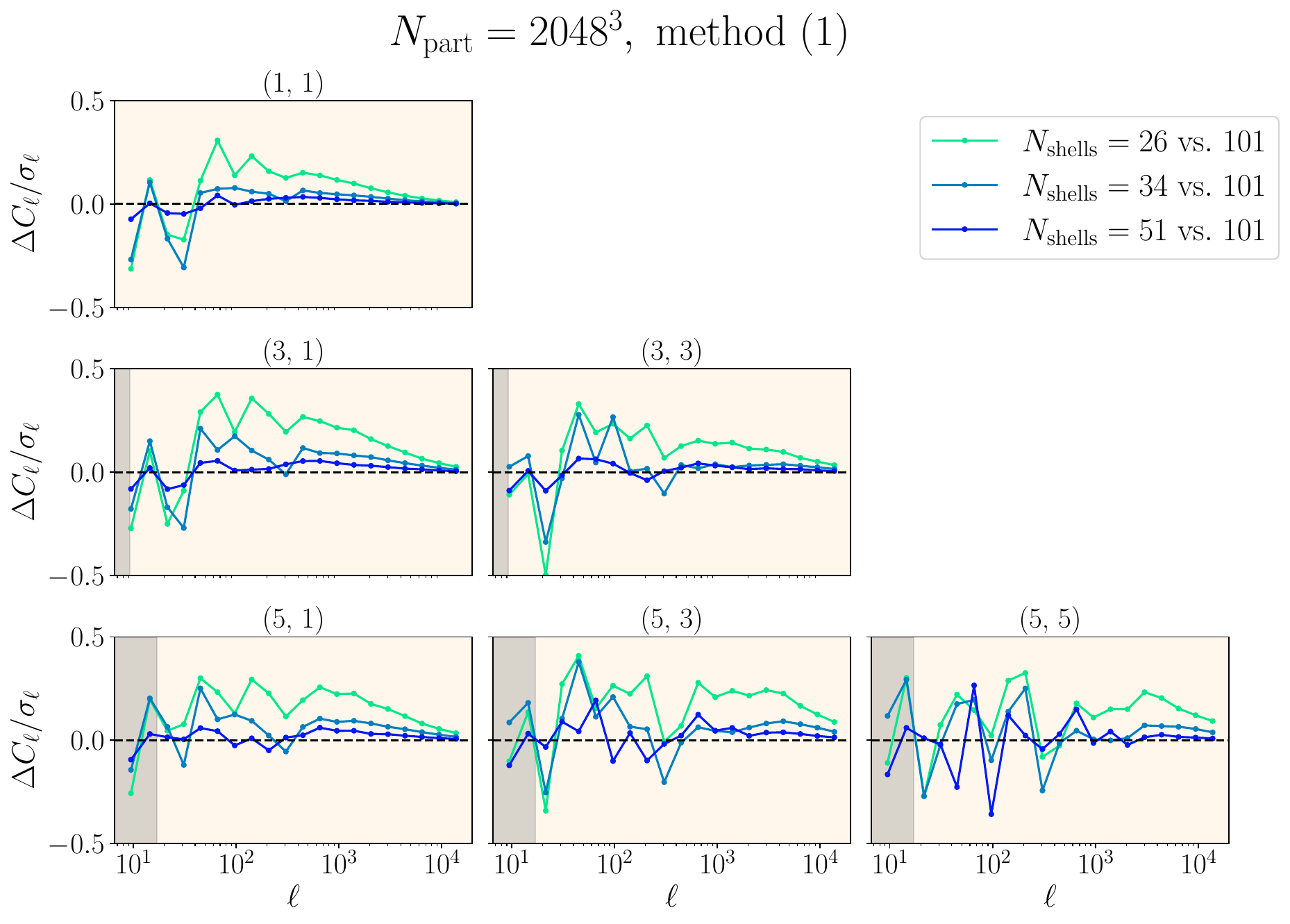}
    \caption{Convergence of the \methodone algorithm (method \ref{item:method1}) with the number of shells. The plot shows $\Delta C_\ell\equiv C_\ell^{\nshells}-C_\ell^{101}$ of the tomographic $\kappa$ maps for $\nshells = 26$, 34 and 51, normalized by the diagonal of the Gaussian covariance matrix (for the first, third and fifth redshift bins). Gray-shaded areas indicate scales larger than the size of the box.}
    \label{fig:Cell_all_method1}
\end{figure*}

In \cref{fig:Cell_all_method1}, we show the $\Delta C_\ell\equiv C_\ell^{\nshells}-C_\ell^{101}$ of the tomographic $\kappa$ maps for $\nshells = 26$, 34 and 51, normalized by the diagonal of the Gaussian covariance matrix. The figure demonstrates that, while any differences in the measured angular power spectra are below $0.3\sigma$, our results converge to the highest-resolution case ($\nshells = 101$), reaching below $0.1\sigma$ when using 51 snapshots (being a bit noisier in the highest redshift bin). To quantify this statement, in \cref{tab:chi2_method1}, we show the $\chi^2$ calculated from the pairwise comparison of the $C_\ell$ of the convergence maps for the $\nshells = 26$, 34, 51 and 101 cases, computed using \cref{eq:chi2} (corresponding to 286 degrees of freedom). As \cref{fig:Cell_all_method1} suggests, the pairwise $\chi^2$ with respect to 101 decreases as the number of snapshots increases. 

\begin{table}
    \renewcommand{\arraystretch}{1.3} 
    \setlength{\tabcolsep}{7pt} 
    \centering
    \begin{tabular}{ccccc}
        \toprule\toprule
        $\boldsymbol{\nshells}$ & \textbf{26} & \textbf{34} & \textbf{51} & \textbf{101} \\ \hline
        \textbf{26}  & $--$ & 2.0 (1.2) & 4.0 (2.4) & 5.9 (2.8) \\ \hline
        \textbf{34}  & $--$ & $--$ & 1.5 (1.0) & 2.7 (1.8) \\ \hline
        \textbf{51}  & $--$ & $--$ & $--$ & 1.2 (0.7) \\
        \bottomrule\bottomrule
    \end{tabular}
    \caption{Convergence of the \methodone algorithm (method \ref{item:method1}) with the number of shells for $\npart = 2048^3$ (and for $\npart = 1024^3$ in parenthesis). The table shows the matrix of pairwise $\chi^2$ values of the $C_\ell$ between different \nshells cases: 26, 34, 51 and 101. The $\chi^2$ were computed using \cref{eq:chi2} and correspond to 286 degrees of freedom, equal to the total number of multipoles used across the 5 redshift bins.}
    \label{tab:chi2_method1}
\end{table}

\paragraph{Higher-order moments of $\kappa$} 

\begin{figure*}
    \centering
    \includegraphics[width=0.95\textwidth]{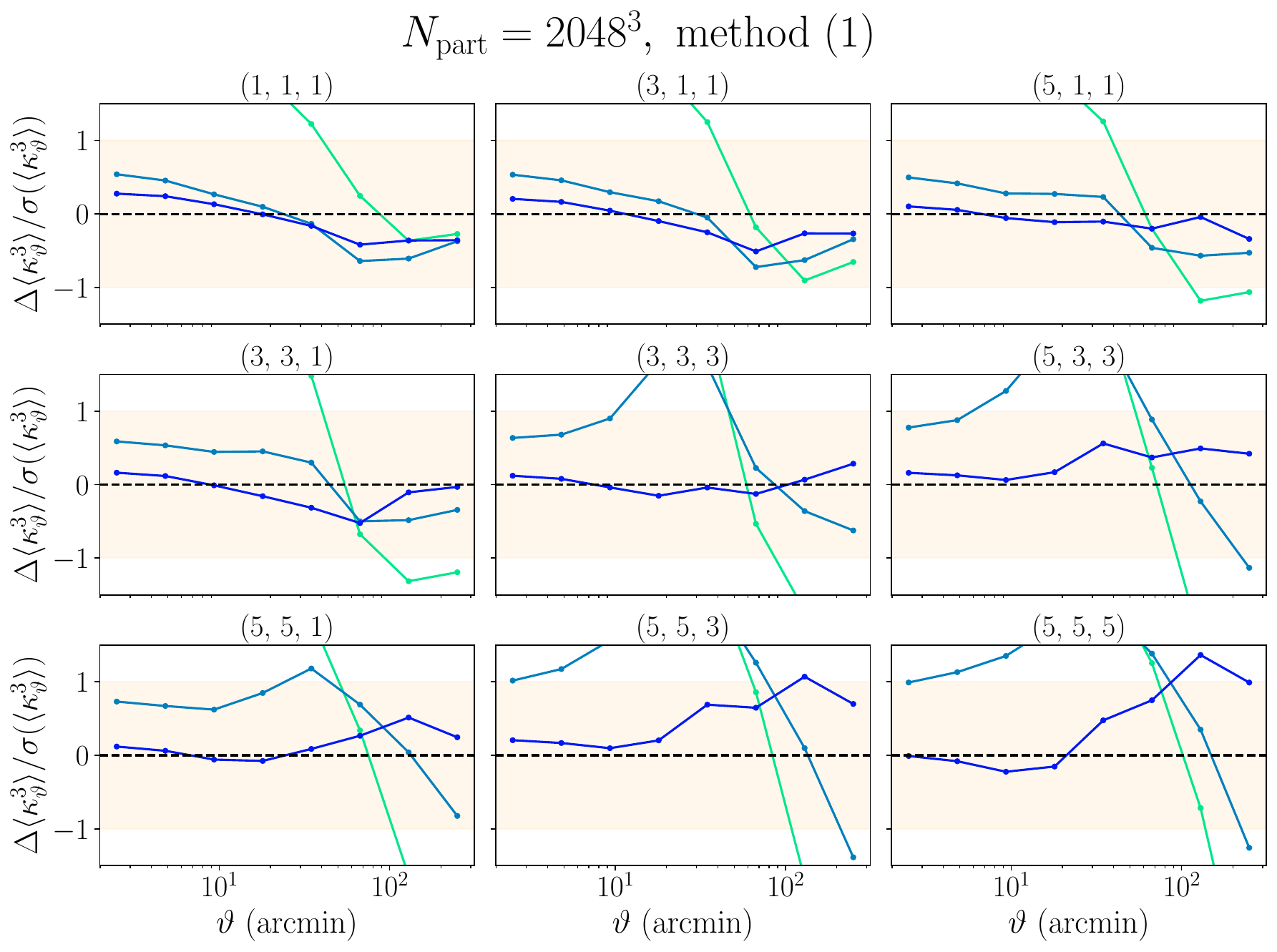}
    \includegraphics[width=0.95\textwidth]{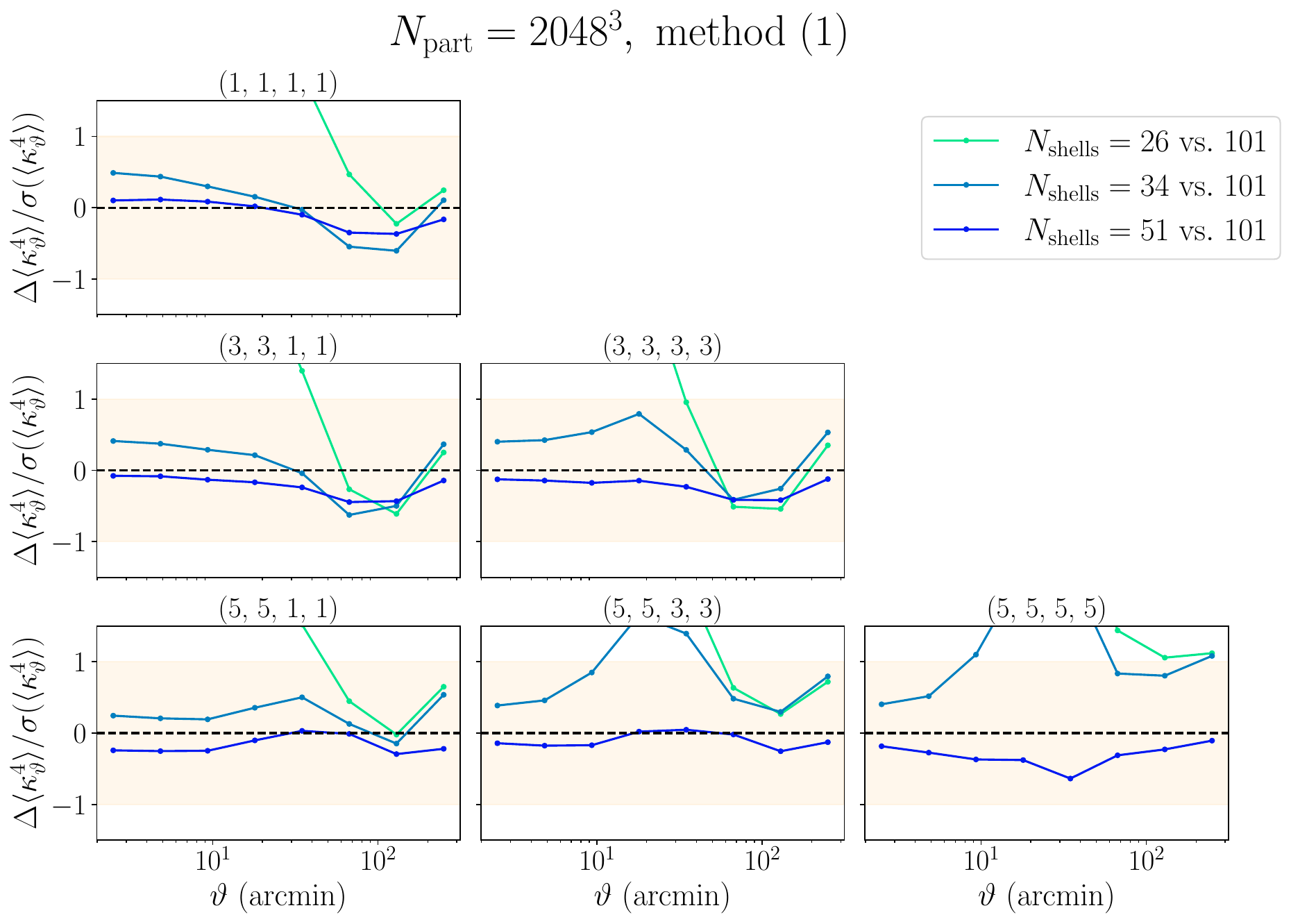}
    \caption{Convergence of the \methodone algorithm (method \ref{item:method1}) with the number of shells. The plot shows $\Delta \langle\kappa^n_\vartheta\rangle\equiv \langle\kappa^n_\vartheta\rangle^{\nshells}-\langle\kappa^n_\vartheta\rangle^{101}$ for $\nshells = 26$, 34 and 51 ($n=3$ on top, $n=4$ on bottom), normalized by the standard deviation of the ten $\nshells = 101$ realizations (for the first, third and fifth redshift bins). The orange-shaded areas indicate the $1\sigma$ regions.}
    \label{fig:kappa_all_method1}
\end{figure*}

In \cref{fig:kappa_all_method1}, we show the same convergence test as in \cref{fig:Cell_all_method1} but for the third (top) and fourth (bottom) moments of $\kappa$.\footnote{We do not show the second moment explicitly, as it conveys similar information to the $C_\ell$. We have verified that it also converges as \nshells increases.} Unlike the $C_\ell$ case, where Gaussian covariance matrices were used in the denominator, the measurements shown here are normalized by the standard deviation estimated from the ten realizations (calculated for $\nshells = 101$). As \nshells increases, the pairwise differences relative to the $\nshells = 101$ result shrink, indicating convergence. In particular, the difference between $\nshells = 26$ and $\nshells = 101$ exceeds $1\sigma$ across nearly all scales and redshift bins, whereas the difference between $\nshells = 51$ and $\nshells = 101$ remains well below $0.5\sigma$ almost everywhere, with the exception of the third moment in the highest-redshift bin, which shows a deviation slightly above $1\sigma$. These results indicate that using $\nshells = 51$ is sufficient to achieve reliable accuracy for the higher-order $\kappa$ moments. 

In \cref{app:kappa_pdf}, we present an analogous analysis for the first-order moment of the $\kappa$ field, the Probability Distribution Function ($\kappa$-PDF). We relegate the $\kappa$-PDF to an appendix because it was less extensively studied during the development of this project and will be explored more thoroughly in a forthcoming work.

The results we have shown so far suggest that our $\npart = 2048^3$ simulations converge as a function of \nshells for both two-point and higher-order statistics, and that using $\nshells = 51$ yields very small differences with respect to the highest-resolution setting (within the $1\sigma$ statistical uncertainty on all scales and redshifts for both two-point and higher-order statistics). We note, however, that slicing snapshots can introduce boundary effects (e.g., halos intersecting shell edges), which limit exact convergence and may have a non-negligible impact on higher-order moments, motivating method comparisons in \cref{sec:results_method2,sec:results_method3}.

\subsubsection{Reducing the number of particles, \npart}\label{sec:results_method1_1024}

Our goal here is to study the accuracy and precision of our simulations when lowering the number of simulation particles, in particular using $\npart = 1024^3$ (instead of $\npart = 2048^3$ as previously). Reducing the number of particles results in a poorer mass resolution, $m_{\rm part} = 2.08\times10^{10}\,M_\odot$, compared to the previous value of $m_{\rm part} = 2.60\times10^9\,M_\odot$.

\paragraph{Comparison with $\npart = 2048^3$}

Here, we directly compare the results of the $\npart = 2048^3$ and $\npart = 1024^3$ simulations. We emphasize that perfect agreement between the two is not expected, since they were generated with different initial conditions. The comparison here is performed using the angular power spectrum rather than higher-order moments of $\kappa$, since the power spectrum can be readily normalized against theoretical predictions.

\begin{figure*}
    \centering
    \includegraphics[width=\linewidth]{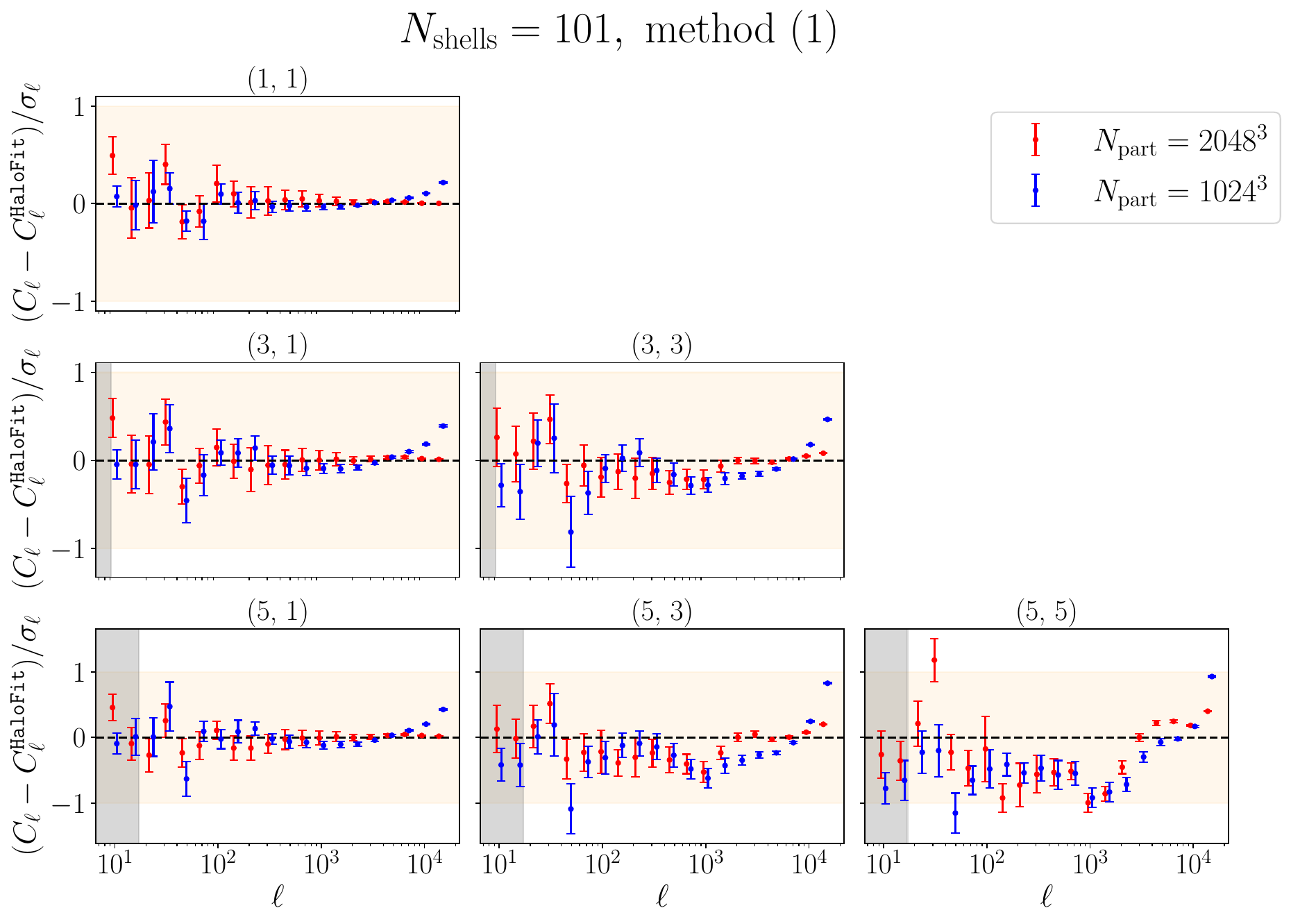}
    \caption{Comparison of the results for $\npart = 2048^3$ vs. $\npart = 1024^3$ for the \methodone algorithm (method \ref{item:method1}). The plot shows the difference in the $C_\ell$ of the tomographic $\kappa$ maps with respect to the prediction from \halofit, normalized by the diagonal of the Gaussian covariance matrix (for the first, third and fifth redshift bins using the SRD redshift distributions). We include the results obtained when constructing the lightcones using $\nshells = 101$, for both $\npart = 2048^3$ (red) and $\npart = 1024^3$ (blue, shifted to higher $\ell$ for visualization purposes). Gray-shaded areas indicate scales larger than the size of the box, and the orange-shaded ones show the $1\sigma$ regions from a Gaussian covariance matrix estimated for the LSST Y10 sample. The error bars were obtained from the variance of the ten realizations. The $\npart = 2048^3$ and $\npart = 1024^3$ simulations use different random seeds, introducing a contribution from cosmic variance.}
    \label{fig:Cell_all_method1_2048_vs_1024}
\end{figure*}

In \cref{fig:Cell_all_method1_2048_vs_1024}, we show the angular power spectrum of the $\kappa$ maps for both $\npart = 2048^3$ and $\npart = 1024^3$, relative to the prediction from \halofit and normalized by the diagonal of the Gaussian covariance (note that we perform a more detailed comparison to theory in \cref{sec:results_theory}). We only include the results for $\nshells = 101$ as an example. We find that on intermediate and large scales ($100 \lesssim \ell \lesssim 5000$), both the $\npart = 2048^3$ (blue) and $\npart = 1024^3$ (red) simulations are in very good agreement with the theoretical predictions. The comparison between the two \npart cases does not show perfect agreement, however, because, as we mentioned earlier, they were generated with different random seeds, introducing a contribution from cosmic variance. At small scales ($\ell \gtrsim 5000$), the $\npart = 1024^3$ simulation exhibits some excess power, which is due to particle shot noise.

\paragraph{Convergence with \nshells for $\npart = 1024^3$}

Here we study the convergence of our $\npart = 1024^3$ simulations as a function of \nshells for both the $C_\ell$ and higher-order moments of $\kappa$.
\begin{itemize}
    \item Angular power spectrum. An analogous plot of \cref{fig:Cell_all_method1} but for $\npart = 1024^3$ is shown in \cref{fig:Cell_all_method1_1024} of \cref{app:1024}, which suggests our results converge as \nshells increases for the $\npart = 1024^3$ case. The $\chi^2$ obtained are listed in \cref{tab:chi2_method1} (in parentheses). We find that the $\chi^2$ decreases as \nshells increases, and therefore we conclude that $\npart = 1024^3$ seems to be sufficient for the convergence of two-point statistics.
    
    \item Higher-order moments of $\kappa$. An analogous plot of \cref{fig:kappa_all_method1} but for $\npart = 1024^3$ is shown in \cref{fig:kappa_all_method1_1024} of \cref{app:1024}, where we show the third and fourth moments of the $\kappa$ field. Unlike in the $\npart = 2048^3$ case, we do not see a clear convergence with \nshells (the residuals are of the order of $1\sigma$), especially for the fourth-order moments (akin to the trispectrum). For this reason, we conclude that the mass resolution $m_{\rm part} = 2.08\times10^{10}\,M_\odot$ obtained with $\npart = 1024^3$ is insufficient to achieve convergence with \nshells for the higher-order moments of $\kappa$. This is in contrast to the $C_\ell$ case, for which convergence was found.
\end{itemize}

\subsection{Comparison with \methodtwo}\label{sec:results_method2}

We now compare the results obtained with \methodtwo (method \ref{item:method2}) to those from \methodone (method \ref{item:method1}).

\begin{figure*}
    \centering
    \includegraphics[width=\linewidth]{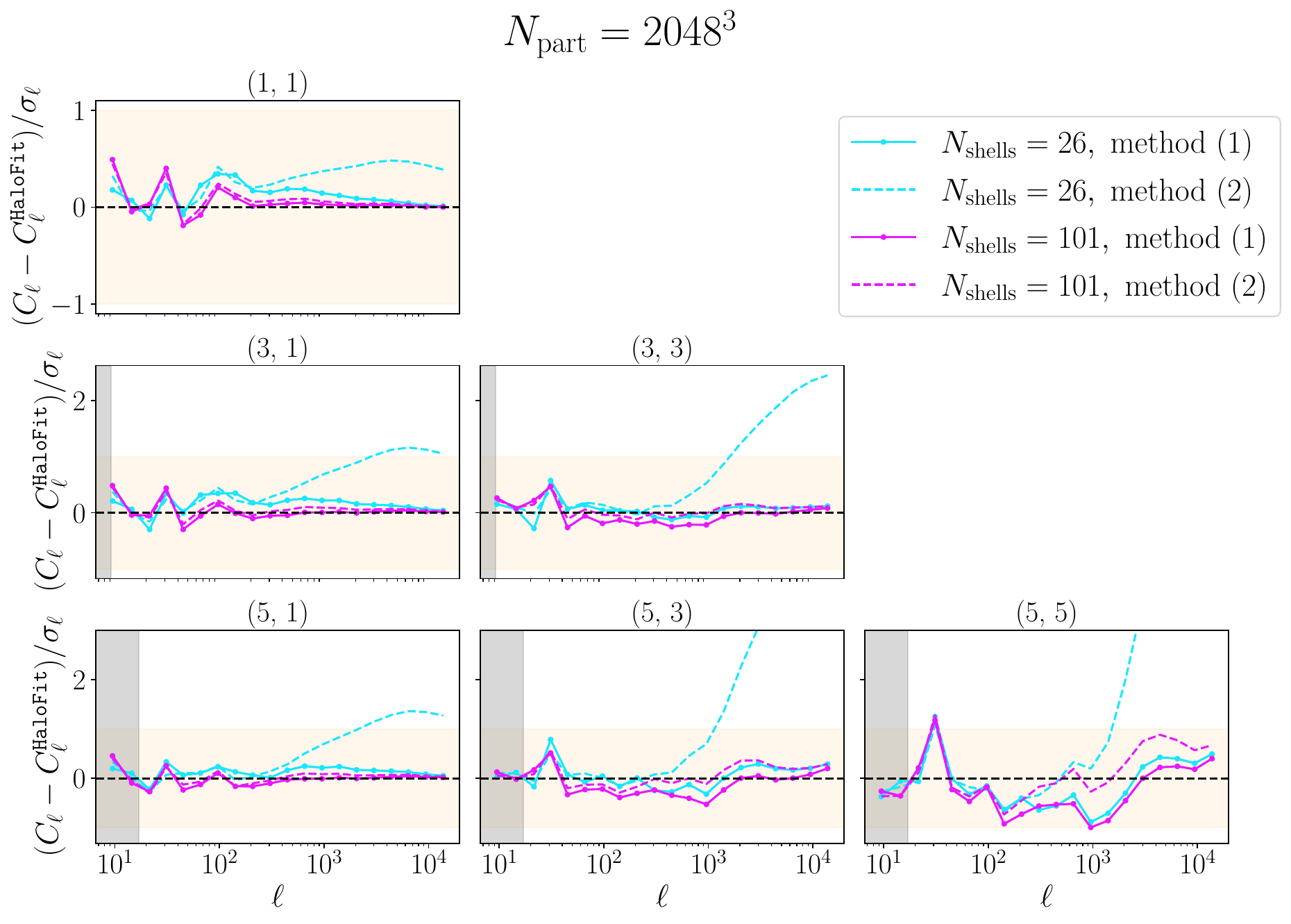}
    \caption{Comparison of the results of the \methodone (method \ref{item:method1}) vs. the \methodtwo (method \ref{item:method2}) algorithms. The plot shows the difference in the $C_\ell$ of the tomographic $\kappa$ maps with respect to the prediction from \halofit, normalized by the diagonal of the Gaussian covariance matrix (for the first, third and fifth redshift bins using the SRD redshift distributions). We include the results obtained when constructing the lightcones using $\nshells = 26$ (cyan) and 101 (fuchsia), for both methods \ref{item:method1} (solid) and \ref{item:method2} (dashed). Gray-shaded areas indicate scales larger than the size of the box, and the orange-shaded ones show the $1\sigma$ regions from a Gaussian covariance matrix estimated for the LSST Y10 sample.}
    \label{fig:method1_vs_method2}
\end{figure*}

\Cref{fig:method1_vs_method2} shows the $C_\ell$ of the tomographic $\kappa$ maps for both methods, for $\nshells = 26$ (cyan) and 101 (fuchsia). At fixed \nshells, method \ref{item:method2} exhibits an excess of small-scale power relative to method \ref{item:method1}. This effect becomes relevant on progressively smaller angular scales as redshift increases (appearing at $\ell \gtrsim 200$ at low redshift and at $\ell \gtrsim 1000$ at high redshift), and its amplitude also grows with increasing redshift. However, it becomes less noticeable when using more, and therefore thinner, shells: for $\nshells = 26$, the cyan dashed line deviates noticeably from the cyan solid line, whereas the discrepancy is much reduced for $\nshells = 101$, although still present. This is an expected result, since the interpolation of particle worldlines to the lightcone inherently neglects their curved (accelerated) motion and the resulting misplacements inject spurious small-scale fluctuations into the field, enhancing the high-$k$ power (see, e.g., \cite{smith2022light,smith2022solving}).

Aside from this additional small-scale power, the two methods agree well. Although method \ref{item:method2} incorporates power-spectrum evolution along the line of sight, the resulting amplification of small-scale power is detrimental for the statistics considered here. We, therefore, do not consider method \ref{item:method2} further in this work.

\subsection{Comparison with \methodthree}\label{sec:results_method3}

Here, we analyze the results obtained using the \methodthree algorithm (method \ref{item:method3}). Method \ref{item:method3} is expected to give more accurate results compared to method \ref{item:method1} since its time resolution is inherently higher, since it includes sub-shell evolution. Our goal in this section is to validate the results of method \ref{item:method1}.

For method \ref{item:method3}, we construct lightcones with different numbers of shells while always using all available \lightconeoutputs.\footnote{In method \ref{item:method3}, the number of \lightconeoutputs assigned to each shell depends on the shell’s width; however, we always use the full set of 400 \lightconeoutputs to build the lightcones (unlike in methods \ref{item:method1} and \ref{item:method2}, where we use one snapshot output per shell). For example, when constructing a lightcone with $\nshells = 200$, we use two \lightconeoutputs per shell, effectively capturing sub-shell evolution.} The tested configurations are $\nshells = 26$, 34, 51, 101, 201 and 400. All the results here
\begin{enumerate}
    \item use $N$-body simulations with $\npart = 1024^3$ (we used $\npart = 2048^3$ for both methods \ref{item:method1} and \ref{item:method2}).
    \item are averaged over two simulation seeds and a single random seed for the observers, i.e., averaged over a total of two data vectors (we used five random seeds for the observers per simulation seed for both methods \ref{item:method1} and \ref{item:method2}). The reason is that HACC implements a more traditional approach to rotate and replicate the boxes as detailed in \cref{sec:algorithms}, which only allows one to create a single lightcone per $N$-body simulation.
    \item only include particles up to redshift 1 (we extended our analysis to redshift 4 for both methods \ref{item:method1} and \ref{item:method2}). We still use the SRD redshift distributions shown in \cref{fig:nz_qz_SRD}, but limited to the first three redshift bins. The results obtained in this section for $0 < z < 1$ are expected to remain valid at higher redshifts.
\end{enumerate}

\begin{figure*}
    \centering
    \includegraphics[width=\textwidth]{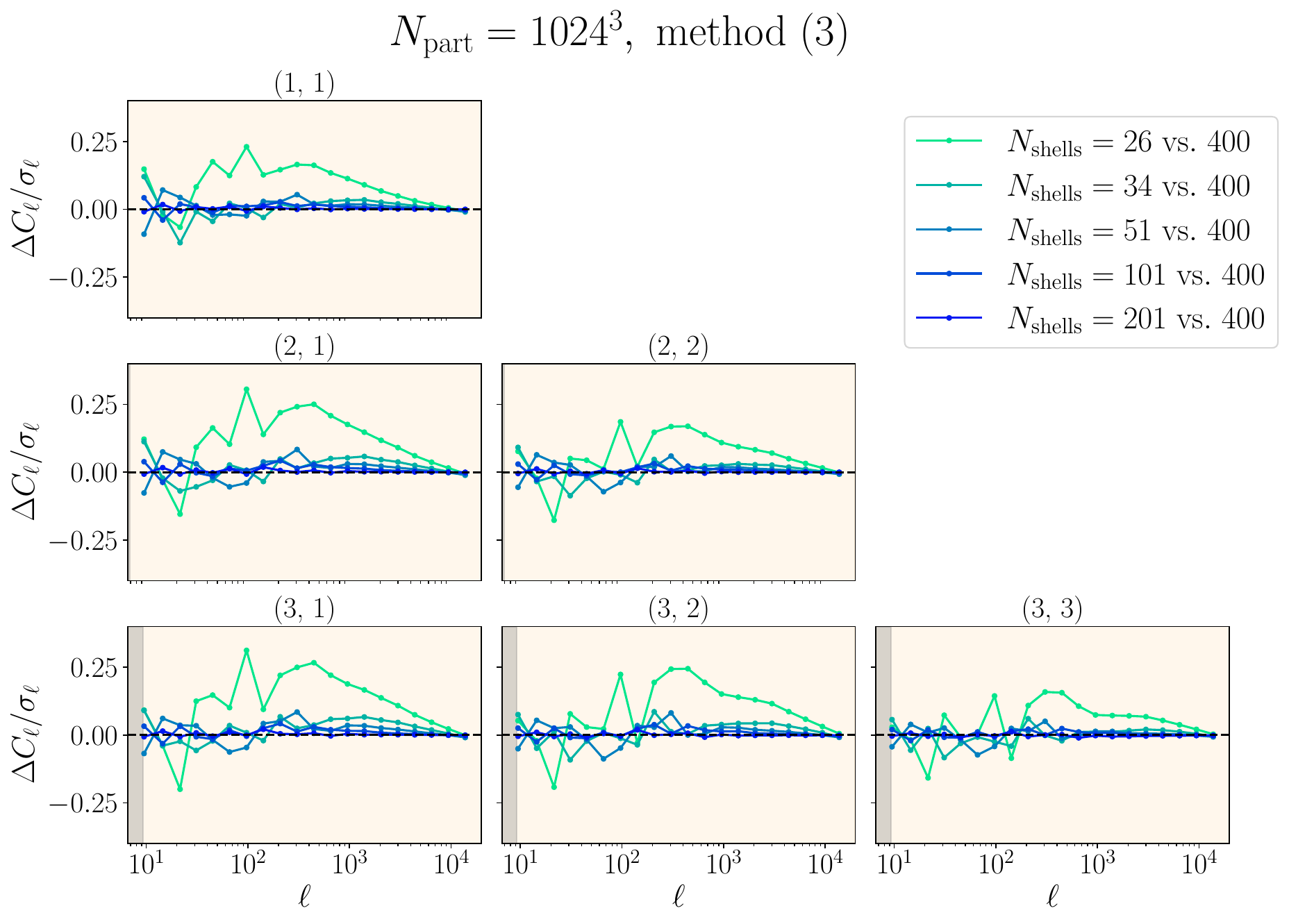}
    \caption{Convergence of the \methodthree algorithm (method \ref{item:method3}) with the number of shells. The plot shows $\Delta C_\ell\equiv C_\ell^{\nshells}-C_\ell^{400}$ of the tomographic $\kappa$ maps for $\nshells = 26$, 34, 51, 101 and 201, normalized by the diagonal of the Gaussian covariance matrix (for the first three redshift bins). Gray-shaded areas indicate scales larger than the size of the box.}
    \label{fig:Cell_all_4_method3}
\end{figure*}

In \cref{fig:Cell_all_4_method3} and \cref{tab:chi2_method3}, we show the pairwise $C_\ell$ and report the associated pairwise $\chi^2$ values (corresponding to 120 degrees of freedom), respectively, for all combinations of $\nshells$ obtained using method \ref{item:method3}. As \nshells increases, the $C_\ell$ converge, reflected by the decreasing $\chi^2$. In particular, the $\chi^2$ relative to $\nshells = 400$ is below 0.10 for $\nshells \geq 51$. This suggests that the sub-shell evolution captured by method \ref{item:method3}, but not by method \ref{item:method1}, can be neglected for $\nshells \geq 51$. We also note that these $\chi^2$ values are significantly smaller than those in \cref{tab:chi2_method1}. This is partly because here the analysis is restricted to the first three redshift bins (and, therefore, the number of data points is 42\% smaller), and partly because here the effect of \nshells is limited to discretization, while the evolution of the power spectrum is naturally included (unlike in methods \ref{item:method1} and \ref{item:method2}).

\begin{table}
    \renewcommand{\arraystretch}{1.3} 
    \setlength{\tabcolsep}{7pt} 
    \centering
    \begin{tabular}{ccccccc}
        \toprule\toprule
        $\boldsymbol{\nshells}$ & \textbf{26} & \textbf{34} & \textbf{51} & \textbf{101} & \textbf{201} & \textbf{400} \\ \hline
        \textbf{26}  & $--$  & 1.05  & 1.18  & 1.14  & 1.28  & 1.29  \\ \hline
        \textbf{34}  & $--$  & $--$  & 0.22  & 0.11  & 0.15  & 0.15  \\ \hline
        \textbf{51}  & $--$  & $--$  & $--$  & 0.09  & 0.07  & 0.08  \\ \hline
        \textbf{101} & $--$  & $--$  & $--$  & $--$  & 0.02  & 0.02  \\ \hline
        \textbf{201} & $--$  & $--$  & $--$  & $--$  & $--$  & 0.00  \\
        \bottomrule\bottomrule
    \end{tabular}
    \caption{Convergence of the \methodthree algorithm (method \ref{item:method3}) with the number of shells. The table shows the matrix of pairwise $\chi^2$ values of the $C_\ell$ between different \nshells cases: 26, 34, 51, 101, 201 and 400. The $\chi^2$ values correspond to 120 degrees of freedom, equal to the total number of multipoles used across the 3 redshift bins.}
    \label{tab:chi2_method3}
\end{table}

\begin{figure*}
    \centering
    \includegraphics[width=\linewidth]{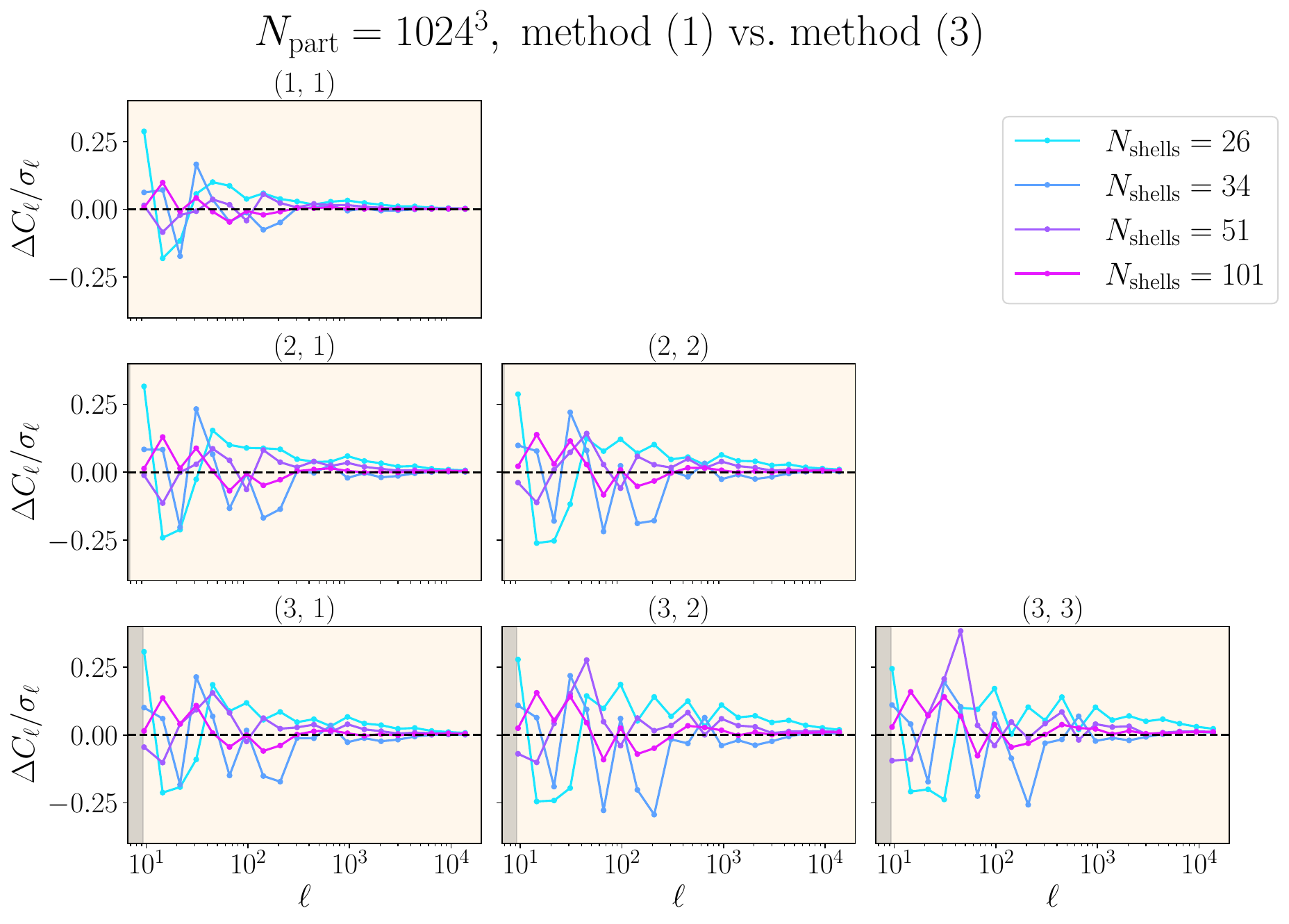}
    \caption{Comparison of the results of the \methodone (method \ref{item:method1}) vs. the \methodthree (method \ref{item:method3}) algorithms. The plot shows $\Delta C_\ell\equiv C_\ell^{\nshells,{\rm method}\ 1}-C_\ell^{\nshells,{\rm method}\ 3} $ of the tomographic $\kappa$ maps for different values of \nshells, normalized by the diagonal of the Gaussian covariance matrix (for the first three redshift bins). Grey-shaded areas indicate scales larger than the size of the box.}
    \label{fig:Cell_all_5_method1_vs_method3}
\end{figure*}

In \cref{fig:Cell_all_5_method1_vs_method3}, we show the pairwise comparison of the angular power spectra obtained with methods \ref{item:method1} and \ref{item:method3} for the common cases of $\nshells = 26$, 34, 51 and 101. We see that both methods agree very well and that their agreement increases with \nshells on all scales. The associated pairwise $\chi^2$ values between methods \ref{item:method1} and \ref{item:method3} are reported in \cref{tab:chi2_method1_vs_method3} (corresponding to 120 degrees of freedom). The $\chi^2$ values in the diagonal of the table are all below 1, showcasing a very good agreement between both methods. We also note that they decrease as \nshells increases, as expected. For $\nshells = 101$, the $\chi^2$ is below 0.10, showcasing an excellent agreement.

\begin{table}
    \centering
    \begin{tabular}{cccccc}
        \toprule\toprule
        & \multirow{2}{*}{$\boldsymbol{\nshells}$} & \multicolumn{4}{c}{\textbf{Method \ref{item:method3}}} \\ 
        \cmidrule(lr){3-6}
        & & \textbf{26} & \textbf{34} & \textbf{51} & \textbf{101} \\ 
        \cmidrule(lr){1-6}
        \multirow{4}{*}{\rotatebox{90}{\textbf{Method \ref{item:method1}}$\ \ \ $}} 
        & \textbf{26}  & 0.57 & 2.31 & 2.73 & 2.56 \\ \cmidrule(lr){2-6}
        & \textbf{34}  & 1.54 & 0.44 & 0.57 & 0.54 \\ \cmidrule(lr){2-6}
        & \textbf{51}  & 1.20 & 0.48 & 0.33 & 0.37 \\ \cmidrule(lr){2-6}
        & \textbf{101} & 1.18 & 0.20 & 0.13 & 0.09 \\
        \bottomrule\bottomrule
    \end{tabular}
    \caption{Comparison of the results for methods \ref{item:method1} and \ref{item:method3}. Pairwise $\chi^2$ values of the $C_\ell$ between methods \ref{item:method1} (rows) and \ref{item:method3} (columns) for different \nshells cases: 26, 34, 51 and 101. The $\chi^2$ values correspond to 120 degrees of freedom, equal to the total number of multipoles used across the 3 redshift bins.}
    \label{tab:chi2_method1_vs_method3}
\end{table}

\subsection{Comparison with theory}\label{sec:results_theory}

The results presented so far have allowed us to confirm the convergence of our simulations with \nshells, i.e., to establish their precision. However, we have not yet studied their accuracy with respect to theoretical predictions. In this section, we compare the $C_\ell$ measured in our simulations with the predictions from \halofit and two different emulators:
\begin{itemize}
    \item \halofit is an empirical fitting function, calibrated on $N$-body simulations, that approximates the non-linear matter power spectrum as a function of the linear one \cite{smith2003halofit, takahashi2012halofit, mead2015halofit}. While it provides a convenient and widely used approach, it is not a purely theoretical prediction and can deviate from simulation results by up to $\sim 10\%$, especially on highly non-linear scales \cite{mead2021hmcode}.

    \item The Mira-Titan IV emulator (MTIV), developed by \cite{moran2023mira}, is a state-of-the-art tool designed to predict nonlinear matter power spectra across a broad range of cosmological parameters. It builds upon the Mira-Titan simulation suite, an ensemble of 111 high-resolution $N$-body simulations, each covering a $(2.1\Gpc)^3$ volume and evolving $3200^3$ particles. These simulations span an extensive 6-dimensional cosmological parameter space $\{\Omega_{\rm m}, \Omega_{\rm b}, h, \sigma_8, n_{\rm s}, w\}$ and, in addition, include massive neutrinos and an equation of state for dynamical dark energy. MTIV employs Gaussian process regression techniques to interpolate power spectra between these simulation points, achieving percent-level accuracy over scales down to $k\sim5\iMpc$ and redshifts $z\lesssim 2$.

    \item The \textit{Euclid} Emulator 2 (EE2), presented by \cite{euclid2021euclid}, is built on an extensive set of $N$-body simulations covering an 8-dimensional cosmology space $w_0w_a$CDM+$\sum m_\nu$. It is built on an ensemble of 250 high-resolution simulations, each covering a volume of $(1\ihGpc)^3$ and evolving $3000^3$ particles. EE2 achieves subpercent precision in the nonlinear correction emulated for wavenumbers in the range $0.01\hiMpc<k<10\hiMpc$ at redshifts between 0 and 3.
\end{itemize}

We compare the results on the simulations with the theoretical predictions of these two emulators at three different levels: (1) at the level of the matter power spectrum measured from the cubic boxes at different redshift snapshots, (2) at the level of the angular power spectrum measured on the density maps from the lightcone shells and (3) at the level of the angular power spectrum measured on the convergence maps from the tomographic bins.

\begin{figure*}
    \centering
    \includegraphics[width=0.35\linewidth]{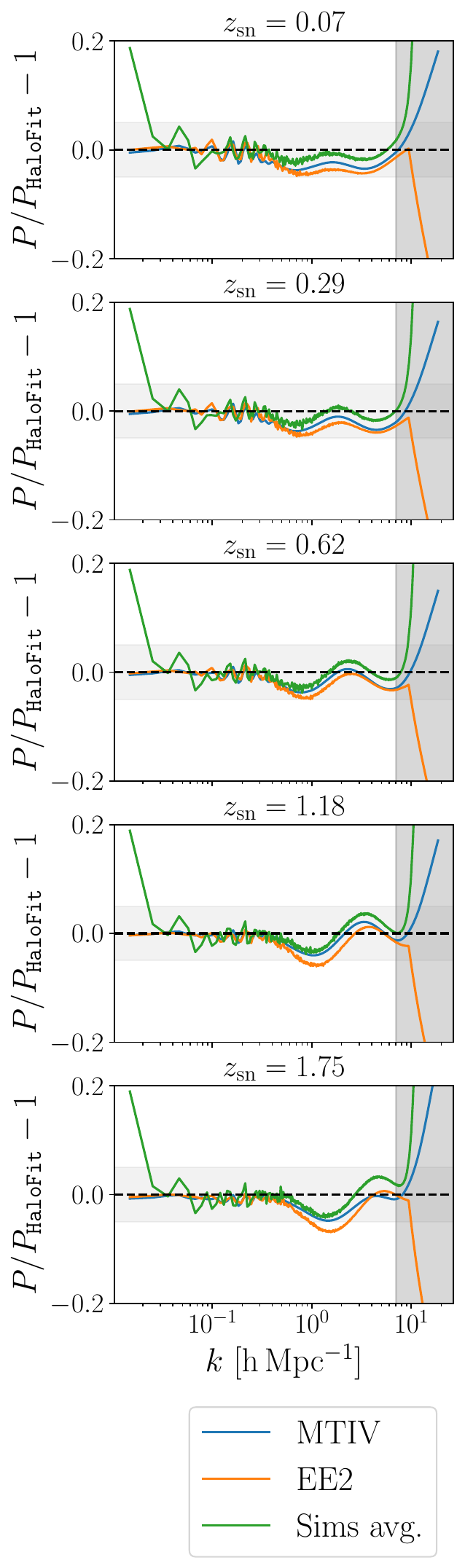}
    \hspace{0.05\linewidth}
    \includegraphics[width=0.353\linewidth]{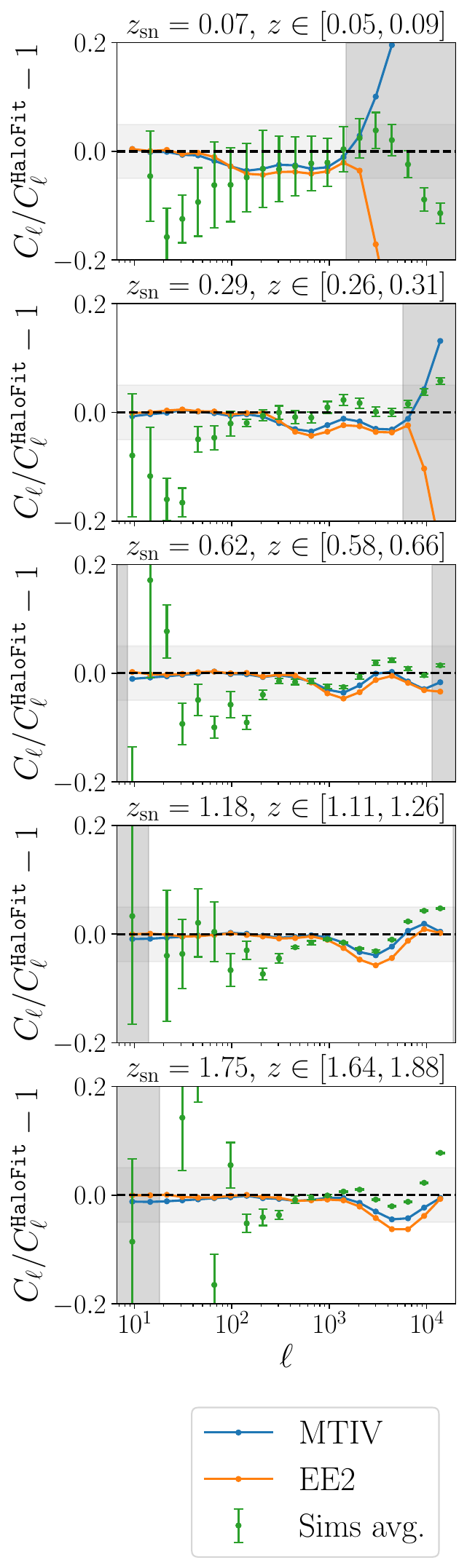}
    \caption{\textbf{Left panel:} $P(k)$ measured on the simulation snapshots (averaged over the two ${\npart = 2048^3}$ simulation seeds) together with that computed with the Mira-Titan IV emulator and the \textit{Euclid} Emulator 2, all of them relative to the prescription from \halofit. The vertical one gray-shaded areas indicate $k>5\iMpc$. \textbf{Right panel:} $C_\ell$ measured on the density shells of the lightcones for $\nshells = 26$ (averaged over the two $\npart = 2048^3$ simulation seeds + five observers per simulation seed, i.e., ten data vectors) together with the predictions from the Mira-Titan IV emulator and the \textit{Euclid} Emulator 2, all of them relative to the prediction from \halofit. The vertical gray-shaded areas indicate $k>5\iMpc$ (when on large $\ell$) or scales larger than the size of the box (when on low $\ell$). In both left and right, we show the measurements for some selected snapshots between redshifts 0 and 4, and the title of each subplot indicates the redshift of the snapshot (and also the redshift range of the shell on the right). The horizontal gray-shaded areas indicate a relative difference of 5\% with respect to \halofit.}
    \label{fig:pk_cell_sims_shells}
\end{figure*}

\begin{figure*}
    \centering
    \includegraphics[width=\linewidth]{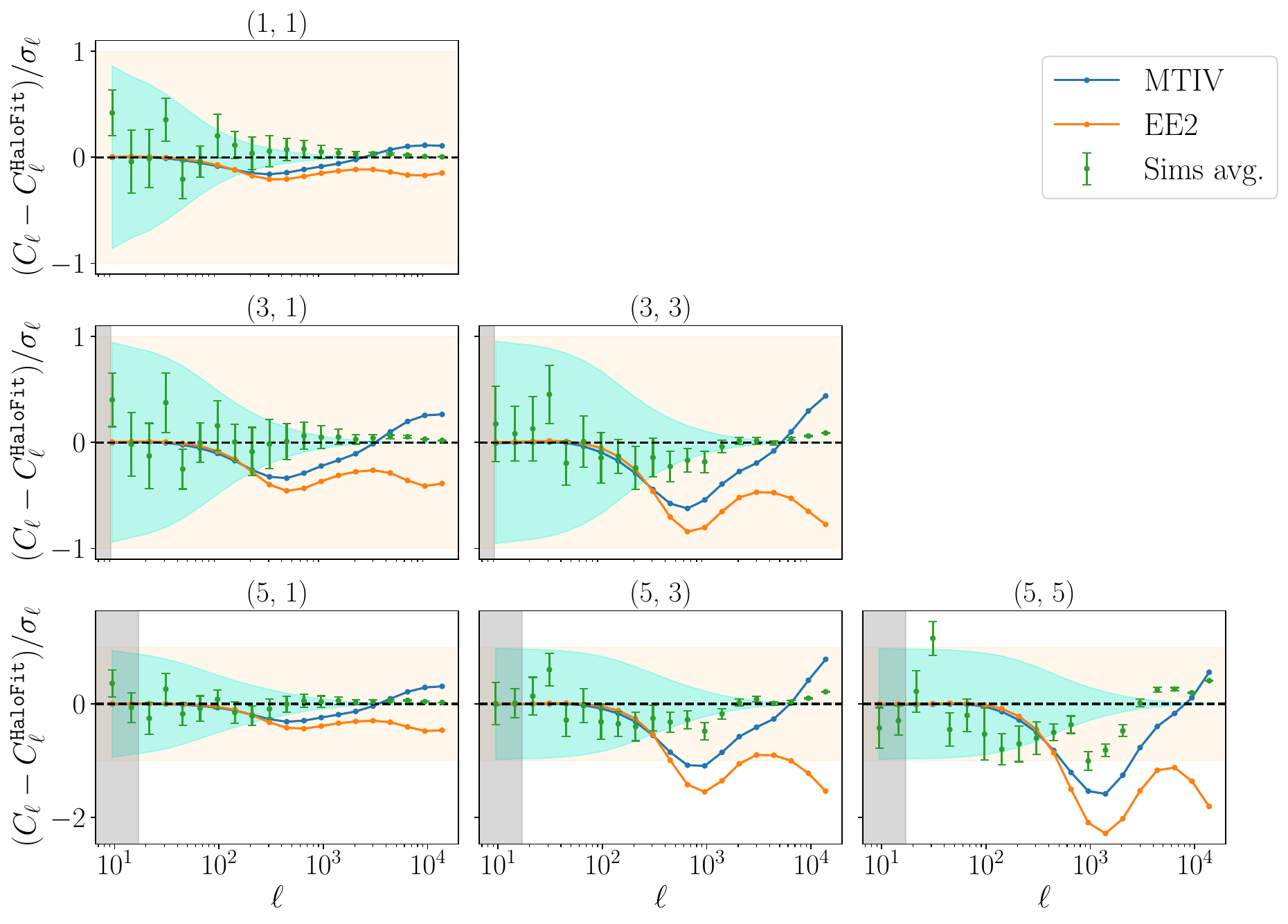}
    \caption{$C_\ell$ measured on the tomographic $\kappa$ maps for $\nshells = 51$ (averaged over the two $\npart = 2048^3$ simulation seeds + five observers per simulation seed, i.e., ten data vectors, green) together with the predictions from the Mira-Titan IV emulator (blue) and the \textit{Euclid} Emulator 2 (orange), all of them relative to the prediction from \halofit. Gray-shaded areas indicate scales larger than the size of the box, the orange-shaded ones show the $1\sigma$ regions and the cyan ones the contribution from cosmic variance (computed using \cref{eq:cov,eq:D_q} assuming $N_q^i=0$). The error bars for the simulations represent the standard deviation across the ten data vectors.}
    \label{fig:Cell_sims_kappa}
\end{figure*}

\begin{enumerate}
    \item $P(k)$ cubic box at different snapshots. In \cref{fig:pk_cell_sims_shells} (left), we show the average power spectrum of the two $\npart = 2048^3$ simulations for some selected snapshots together with the predictions from MTIV and EE2, all of them relative to the prediction from \halofit. These $P(k)$ are measured directly on the simulation snapshots, i.e., on the $N$-body boxes. We observe that the power spectrum of the simulations follows well the predictions of the emulators, particularly the oscillating pattern that appears at $k\gtrsim1\hiMpc$. The simulations agree with the MTIV emulator to within about 3\% over scales where cosmic variance is subdominant ($k \gtrsim 0.05\hiMpc$), up to $k \lesssim 7\hiMpc$. At higher wavenumbers, the emulators extrapolate the power spectrum, and this limit also corresponds, approximately, to the Nyquist frequency of our simulations (see \cref{eq:nyquist} for the $\npart = 2048^3$ case). The agreement holds up to $z \lesssim 2$, beyond which the emulator relies on extrapolation.

    \item $C_\ell$ density maps. In \cref{fig:pk_cell_sims_shells} (right), we show the $C_\ell$ measured in the density shells of our lightcones ($\npart = 2048^3$) for some selected snapshots together with the predictions from MTIV and EE2, projected with the Limber equation using \ccl and including the discretization and in-shell non-evolution of the matter spectrum, as discussed in \cref{sec:discretization}, all of them relative to \halofit. We see that the average of the simulations follows well the trend of MTIV and EE2, better than it follows the prediction from \halofit. Our simulations reach an agreement of $\sim 5\%$ with the MTIV emulator (up to $z\lesssim 2$, since at higher redshifts the emulator extrapolates the power spectrum). Comparing these density $C_\ell$ with the $P(k)$ in the left panel reveals that they exhibit the same oscillating pattern.

    \item $C_\ell$ tomographic $\kappa$ maps. In \cref{fig:Cell_sims_kappa}, we show the $C_\ell$ measured on the tomographic $\kappa$ maps together with the predictions from MTIV and EE2 projected with the Limber equation using \ccl, all of them relative to \halofit. We find that the mean of our simulations ($\nshells = 51$) lies between the MTIV and \halofit predictions, particularly at higher redshift. In the range where the simulations are not dominated by cosmic variance, their deviation from the emulator predictions is comparable to the level of disagreement among the emulators themselves.\footnote{The difference between simulations, \halofit and the emulators cannot be explained by discretization/constant power spectrum, as we expect these effects to be quite small for $\nshells = 51$ (see \cref{fig:binning_2}).} Overall, our simulations remain within $1\sigma$ of the MTIV emulator across all scales.
\end{enumerate}

\section{Conclusions}\label{sec:conclusions}

In this paper, we have systematically investigated the convergence of lightcone simulations for the analysis of weak lensing in the LSST era using higher-order statistics. In particular, we have varied several simulation hyperparameters and lightcone-building methods to test their precision and accuracy. Our analysis employs the LSST SRD source redshift distributions divided into five tomographic bins, assuming Gaussian covariance matrices estimated for the LSST Y10 sample to evaluate the measurement uncertainties.

On the theoretical side, we have shown that sampling linearly in scale factor produces more accurate results than sampling linearly in redshift or comoving distance. We have also quantified the impact of discretizing the lightcone volume into radial shells on the angular power spectrum of the convergence field, as well as the combined effect of discretization and assuming a static power spectrum within each shell. For a configuration with $\nshells = 51$, we find that the resulting biases remain below $0.08\sigma$ compared to the full integral calculation.

For the construction of the lightcones, we have developed \pollux, which we publicly release together with this work (\url{https://github.com/LSSTDESC/pollux}). We have tested three different lightcone-building algorithms: \methodone (method \ref{item:method1}), \methodtwo (method \ref{item:method2}) and \methodthree (method \ref{item:method3}). We have found that methods \ref{item:method1} and \ref{item:method2} agree well on large scales, but the latter produces an expected artificial increase of power on small scales, and therefore method \ref{item:method1} is preferred. Method \ref{item:method3} has an inherently better time resolution compared to method \ref{item:method1}, including sub-shell evolution, but it does not allow us to produce several pseudo-independent realizations from the same $N$-body simulation. Given that method \ref{item:method3} has this major drawback and that it agrees very well on all scales with method \ref{item:method1}, we decided to keep method \ref{item:method1} as our baseline choice for upcoming simulation campaigns.

We have varied the number of shells/snapshots and the number of particles, keeping a fixed comoving size of $L_{\rm box} = 600\ihMpc$. We have generated the lightcones using different numbers of shells between redshifts 0 and 4 (linearly in scale factor), with higher numbers corresponding to thinner shells. We have observed that, in terms of the precision of lensing statistics, our simulations converge as \nshells increases. We conclude that, at least, ${\nshells \gtrsim 50}$ are needed to reach a reasonable precision. As for the density of simulation particles, we have found that $\npart = 1024^3$ (mass resolution of $m_{\rm part} = 2.08\times10^{10}\,M_\odot$) is sufficient to describe two-point statistics, but not for higher-order ones. On the other hand, $\npart = 2048^3$ ($m_{\rm part} = 2.60\times10^9\,M_\odot$) is sufficient to predict both two-point and higher-order statistics, and it is our baseline choice in this work.

We have also concluded that our simulations are consistent with the predictions of the Mira-Titan IV emulator. We have tested this at three different levels: (1) power spectrum of the $N$-body simulation, for which we found that the average $P(k)$ reaches an agreement of $\sim3\%$ with the MTIV emulator on scales between $0.04\hiMpc\lesssim k\lesssim 7\hiMpc$; (2) angular power spectrum of the density maps, for which we found that the average $C_\ell$ reaches an agreement of $\sim5\%$ with the MTIV emulator up to $\ell\lesssim 10^4$; and (3) angular power spectrum of the tomographic convergence maps, for which we found that our simulations are in $1\sigma$ agreement with the predictions from the MTIV emulator on all scales. Finally, the differences between our simulations and the predictions from the emulators are comparable to the differences between the emulators themselves.

On a related note, \pollux can also generate mock galaxy catalogs by sampling the density field and interpolating the corresponding convergence and shear fields at galaxy positions, using the $\delta$, $\kappa$ and $\gamma_{1,2}$ maps produced in this work. Although these catalogs are not used here, they are employed in \cite{bera2025} to study the impact of baryonic effects in a variety of galaxy clustering and weak lensing statistics on small angular scales ($\ell > 3000$). In that analysis, one of our $\nshells = 51$ simulations (produced with method \ref{item:method1}) is baryonified following \cite{schneider2019quantifying} by varying two model parameters, $M_{\rm c}$ and $\theta_{\rm ej}$, which control the extent of gas ejection beyond the halo boundary. The resulting baryonified maps are then processed with \pollux to generate shear and convergence fields and, ultimately, mock catalogs, from which all three types of two-point and four types of three-point correlation functions involving positions and shapes of galaxies are measured and compared to the gravity-only case.

The simulations developed in this paper are also well-suited for the study of observational systematics. In particular, initial efforts have been made to incorporate systematics related to variable survey depth across the footprint. This effect is especially relevant for LSST, whose footprint will not be homogeneous during the first years of observations: some regions of the sky will be observed more frequently and to greater depth than others, leading to spatial variations in source density and redshift distributions. Addressing this effect in a realistic LSST observing scenario is left for future work.

Our final goal is to generate large suites of simulations across a wide space of cosmological parameters, enabling either the construction of a dedicated HOS emulator or allowing us to carry out simulation-based inference analyses. The convergence tests presented here establish the requirements such simulations must meet—most notably in mass resolution and redshift sampling—in order to achieve LSST-level accuracy. The results of this work, therefore, provide a validated foundation for forthcoming large-volume campaigns aimed at fully exploiting the constraining power of cosmic shear using HOS in the LSST era, together with the necessary pipelines for the production of the simulations.

\appendix

\section{Testing the downsampling criterion}\label{app:downsampling}

\begin{figure*}
    \includegraphics[width=\linewidth]{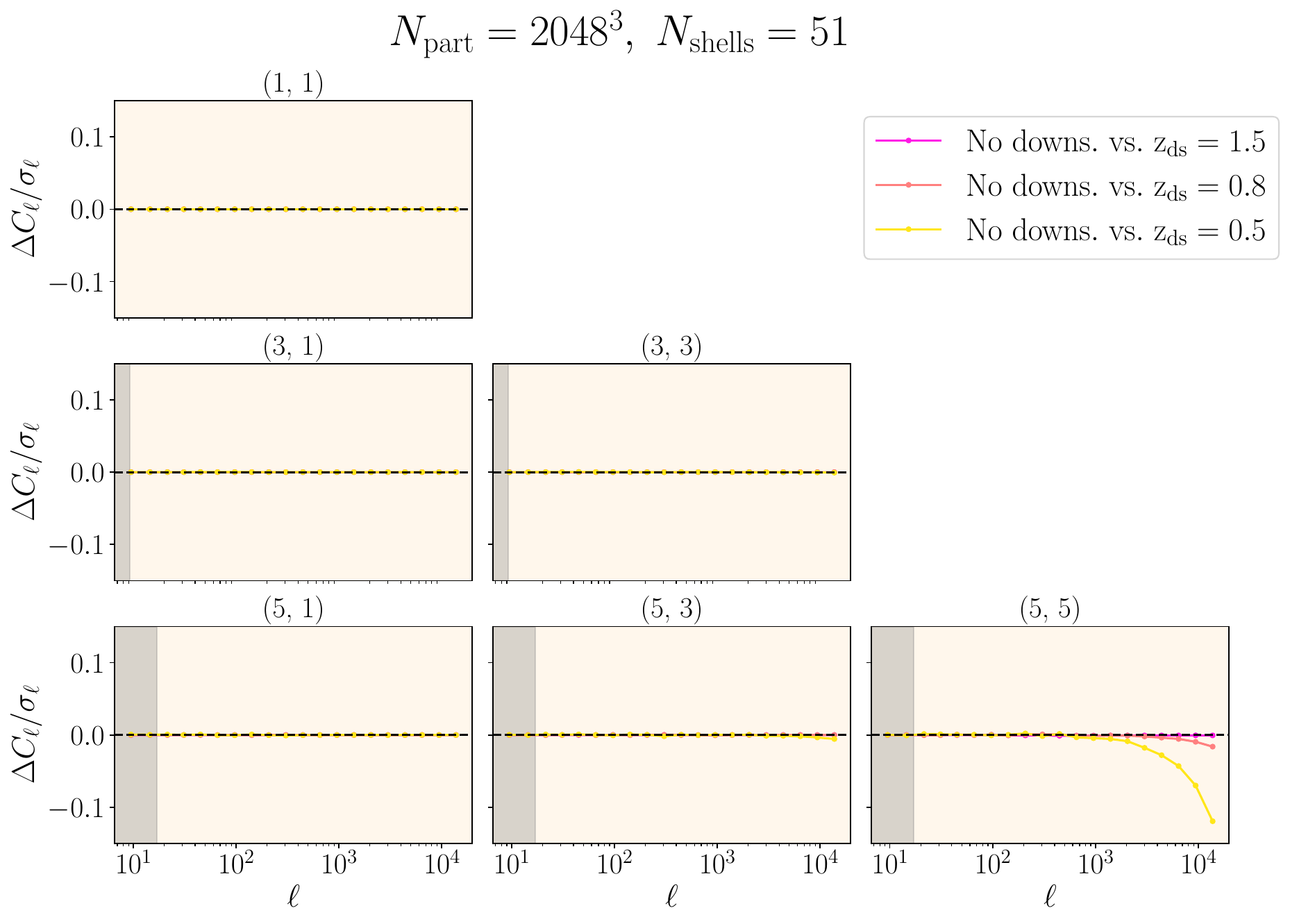}
    \caption{Effect of the downsampling algorithm in the $C_\ell$ of the convergence field. We show the results for $\zdownsampling = 1.5$ (our baseline choice), 0.8 and 0.5, all of them relative to the no-downsample case. Gray-shaded areas indicate scales larger than the size of the box ($\ell_{\rm min}$ is computed using \cref{eq:ellmin}).}
    \label{fig:Cell_all_7_downsampling}
\end{figure*}

In this appendix, we explicitly test the effect of the downsampling criterion described in \cref{sec:pollux} at the level of the angular power spectrum of the convergence field. In \cref{fig:Cell_all_7_downsampling}, we show the differences in the $C_\ell$ with and without downsampling (downsampling using the three values of \zdownsampling mentioned in the main body of the paper: 1.5, 0.8 and 0.5). The downsampling imprints a bias in the $C_\ell$ as \zdownsampling decreases, as expected, and this effect becomes larger with redshift. We find that downsampling from redshift 1.5 produces results indistinguishable from the non-downsampled case, which motivates our choice of $\zdownsampling = 1.5$ as the baseline configuration. This behavior is consistent with the fact that the lensing kernels vanish above $z \approx 1.5$ for all tomographic bins except the highest one, which also rapidly approaches zero beyond that redshift (see \cref{fig:nz_qz_SRD}). We also assessed the impact of downsampling on the third and fourth moments of $\kappa$, finding relative differences below 0.01\% on angular scales between 2.5 and 250 arcmin for the $\zdownsampling = 1.5$ case compared to the non-downsampled case.

\section{\pollux code structure}\label{app:pollux}

\pollux is publicly available at \url{https://github.com/LSSTDESC/pollux}, where example Jupyter notebooks illustrate (i) how to generate density and lensing maps from $N$-body outputs, and (ii) how to read, inspect and visualize the resulting products. The code is organized into a small set of modules that mirror the workflow for constructing lightcone shells, lensing maps and galaxy catalogs:
\begin{itemize}
    \item \texttt{reader.py} — Interfaces with HACC and cubep$^3$m outputs, reads particle/halo data from snapshots or lightcone files and provides the metadata needed to define lightcone shells.
    \item \texttt{map\_builder.py} — Supplies shared \healpix/\healsparse map utilities, spherical-harmonic transforms, angular power spectra computations (\namaster) and common I/O features used by all map-producing components.
    \item \texttt{shells.py} — Implements the \texttt{ShellBuilder}, a child class of the base map builder, which selects particles in each radial slice (with optional subsampling), applies box replications and random isometries and projects particles into HEALPix density maps.
    \item \texttt{lensing.py} — Implements the \texttt{LensingBuilder}, also derived from the base map builder, which stacks density shells under the Born approximation to produce convergence and shear maps, including tomographic combinations.
    \item \texttt{catalog.py} — Generates tomographic galaxy catalogs using uniform or density-biased sampling and interpolates $\kappa$ and $\gamma$ at galaxy positions.
    \item \texttt{cosmo.py} — Defines the cosmology configurations and convenience wrappers for \ccl used by the shell, lensing and catalog builders.
\end{itemize}

\section{\boldmath Convergence of the one-point function with \nshells }\label{app:kappa_pdf}

In recent years, the convergence one-point Probability Distribution Function ($\kappa$-PDF) has emerged as a powerful probe for extracting non-Gaussian information. The $\kappa$-PDF extracts information from different density environments at a given smoothing scale, and it is sensitive to integrals over the whole set of higher-order correlation functions. It offers advantages over moments and cumulants, which capture only partial, and often tail-dominated information about higher-order correlations. In practice, the $\kappa$-PDF offers a description that can be restricted to the bulk of the distribution, which contains non-Gaussian cosmological information and is often more robust to systematics and outliers. It can be easily measured from simulated or observed data of galaxy surveys, and on mildly non-linear scales it can be theoretically predicted with good accuracy \cite{barthelemy2020nulling,boyle2021nuw,barthelemy2024making,castiblanco2024unleashing}. Forecasts on $\sigma_8$ and $w_0$ within the \textit{Euclid} collaboration have shown that constraints from the $\kappa$-PDF alone outperform the two-point correlation functions, and that combining both statistics tightens contours \cite{mellier2024euclid,vinciguerra2025euclid}.

\begin{figure}
    \centering
    \includegraphics[width=\textwidth]{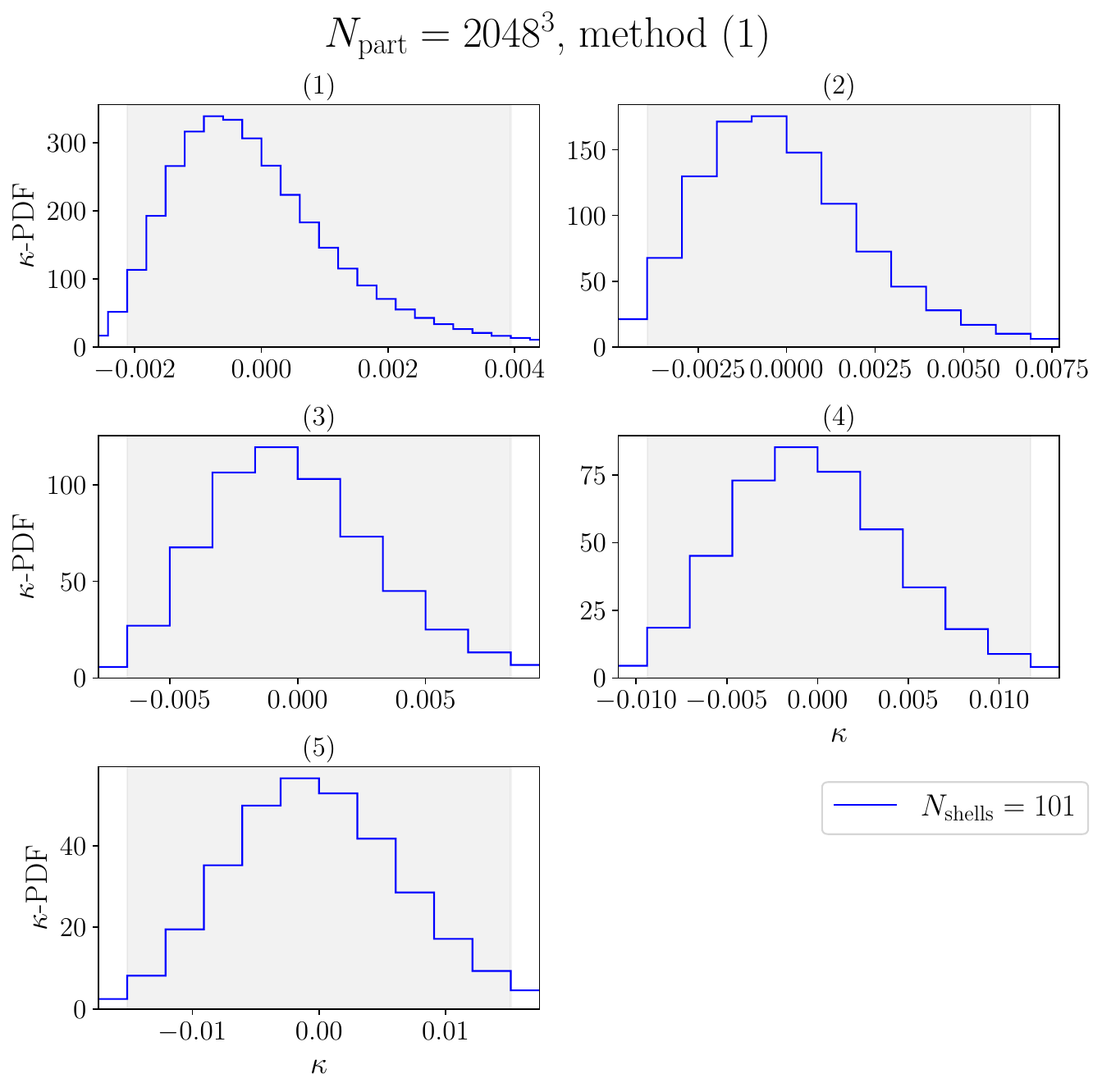}
    \caption{$\kappa-$PDF of the tomographic $\kappa$ maps created with \methodone algorithm (method \ref{item:method1}) for $\nshells=101$ and \nside = 8192. Grey-shaded regions show the intervals containing 95\% of the area of the PDF. }
    \label{fig:kappa_PDF_method1}
\end{figure}

\begin{figure}
    \centering
    \includegraphics[width=\textwidth]{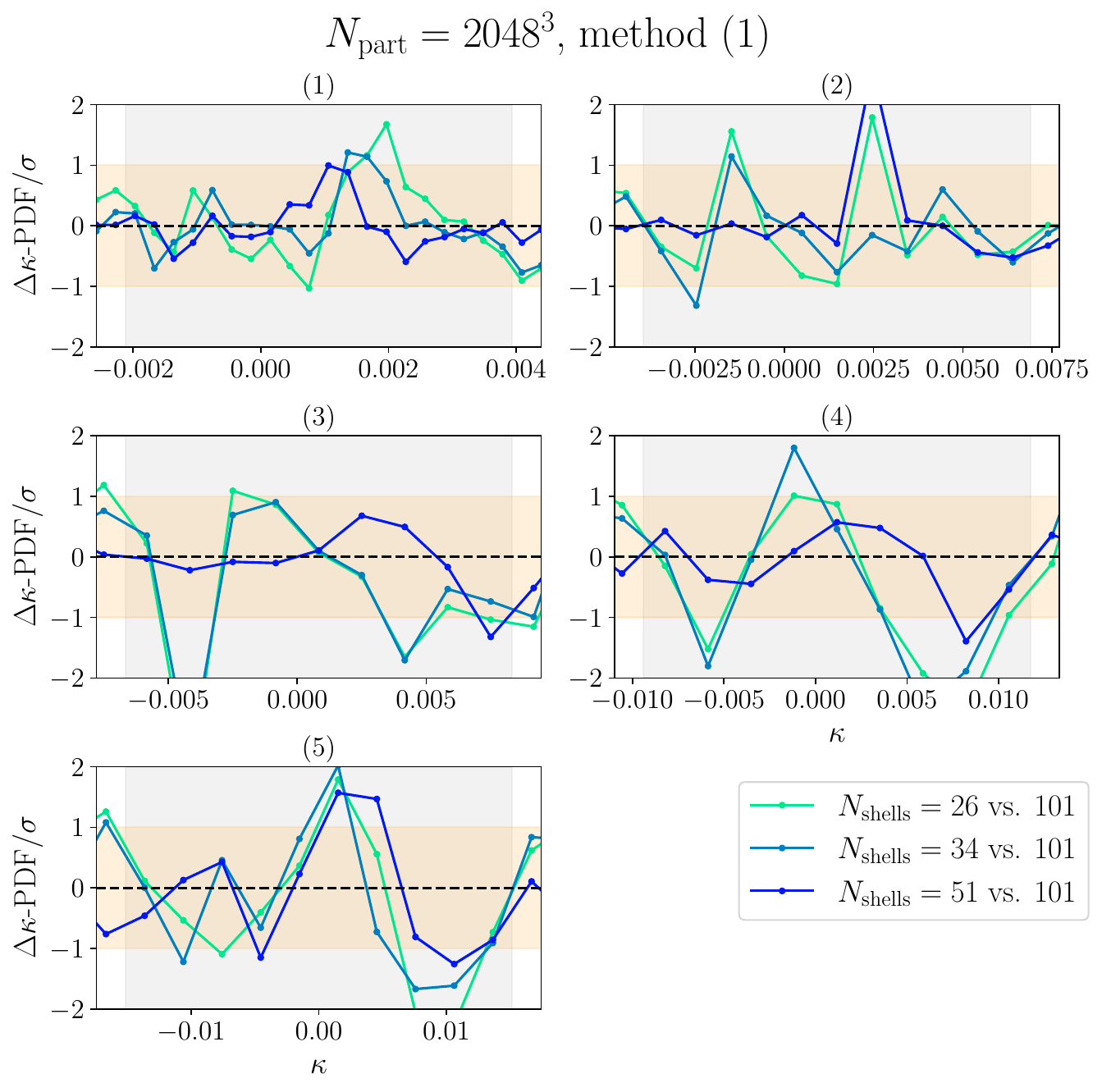}
    \caption{Convergence of the \methodone algorithm (method \ref{item:method1}) with the number of shells. The plot shows $\Delta \mathrm{PDF}\equiv \mathrm{PDF}^{\nshells}-\mathrm{PDF}^{101}$ of the tomographic $\kappa$ maps for $\nshells = 26$, 34 and 51, normalized by the standard deviation of the ten $\nshells = 101$ realizations. The orange-shaded areas indicate the $1\sigma$ regions. Grey-shaded regions show the intervals containing 95\% of the area of the PDF.}
    \label{fig:diff_kappa_PDF_method1}
\end{figure}

From our simulations, we measure the $\kappa$-PDF by constructing a histogram of the pixel values (with a resolution of 0.43 arcmin or \nside = 8192). We adopt a Top-Hat filter with a smoothing scale of 5 arcmin. We choose 100 linearly spaced $\kappa$-bins, with ranges based on the minimum and maximum $\kappa$ values optimized for each tomographic bin. For every bin, we average the $\kappa$-PDF over ten realizations (two simulation seeds + five observers) while keeping the number of shells fixed. The resulting $\kappa$-PDFs for $\nshells = 101$ and all tomographic bins are shown in \cref{fig:kappa_PDF_method1}, where we observe that all PDFs display clear non-Gaussian features, including asymmetric shapes and extended tails. We also see that the variance increases with redshift, as expected from its direct connection with the amplitude of the lensing power spectrum \citep{castiblanco2024unleashing}. To assess the convergence of the simulations generated with the \methodone algorithm (method \ref{item:method1}), we compute the difference in the $\kappa$-PDF between simulations with $\nshells = 26$, 34 and 51, and the reference simulation with $\nshells = 101$. As shown in \cref{fig:diff_kappa_PDF_method1}, the discrepancies systematically decrease as the number of shells increases, with the largest differences appearing in the lowest tomographic bins, as expected from the fewer lens planes available in this portion of the lightcone. We also note that none of these measurements include shape noise, which in fact would smooth out differences even more.

To further quantify this convergence and the shape of the differences, we analyze how the variance, $\sigma_{\kappa}^2 = \langle \kappa^2 \rangle$, and the skewness, $\langle \kappa^3 \rangle / \sigma_{\kappa}^3$, of the $\kappa$-PDF evolve with $\nshells$. The corresponding values are summarized in \cref{tab:PDF_variance} and \cref{tab:PDF_skewness}, respectively. We find that the variance is remarkably stable, with relative differences below $0.6\%$ across all tomographic bins when comparing $\nshells = 26$, 34, 51 and 101. In contrast, the skewness shows a clearer dependence on the number of shells, decreasing monotonically as $\nshells$ increases. However, this effect remains small: the skewness for $\nshells = 26$ differs from the $\nshells = 101$ case by $7\%$, and the difference drops below $0.1\%$ for $\nshells = 51$. This dominant change in the skewness explains the asymmetric behavior of the differences seen in \cref{fig:diff_kappa_PDF_method1}. This quantitative trend is consistent with the behavior observed in the main text for the third and fourth $\kappa$ moments. Overall, these results confirm that simulations generated with $\nshells = 51$ already provide a sufficiently converged estimate of both the $\kappa$ moments and the full $\kappa$-PDF, making them an efficient and reliable choice for our purposes.

\begin{table}
    \centering
    \begin{tabular}{c|rrrr}
        \toprule
        \toprule
        \multirow{2}{*}{\textbf{Tomographic bin}}
            & \multicolumn{4}{c}{\boldmath\nshells} \\
        \cmidrule(lr){2-5}
            & \multicolumn{1}{c}{26}
            & \multicolumn{1}{c}{34}
            & \multicolumn{1}{c}{51}
            & \multicolumn{1}{c}{101} \\
        \midrule
        1 & 2.445 & 2.424 & 2.422 & 2.419 \\
        2 & 6.793 & 6.733 & 6.720 & 6.722 \\
        3 & 13.293 & 13.231 & 13.177 & 13.178 \\
        4 & 24.275 & 24.198 & 24.094 & 24.100 \\
        5 & 52.189 & 52.056 & 51.774 & 51.899 \\
        \bottomrule
        \bottomrule
    \end{tabular}
    \caption{Variance of the $\kappa$-PDF ($\times10^{-6}$) for different \nshells for each tomographic bin.}
    \label{tab:PDF_variance}
\end{table}

\begin{table}
    \centering
    \begin{tabular}{c|rrrr}
        \toprule
        \toprule
        \multirow{2}{*}{\textbf{Tomographic bin}}
            & \multicolumn{4}{c}{\boldmath\nshells} \\
        \cmidrule(lr){2-5}
            & \multicolumn{1}{c}{26}
            & \multicolumn{1}{c}{34}
            & \multicolumn{1}{c}{51}
            & \multicolumn{1}{c}{101} \\
        \midrule
        1 & 1.979 & 1.964 & 1.956 & 1.938 \\
        2 & 1.335 & 1.257 & 1.246 & 1.247 \\
        3 & 0.873 & 0.829 & 0.814 & 0.815 \\
        4 & 0.587 & 0.569 & 0.553 & 0.553 \\
        5 & 0.355 & 0.351 & 0.341 & 0.342 \\
        \bottomrule
        \bottomrule
    \end{tabular}
    \caption{Skewness of the $\kappa$-PDF for different choices for different \nshells for each tomographic bin.}
    \label{tab:PDF_skewness}
\end{table}

\section{\boldmath Convergence with \nshells for $\npart = 1024^3$}\label{app:1024}

\begin{figure*}
    \centering
    \includegraphics[width=\textwidth]{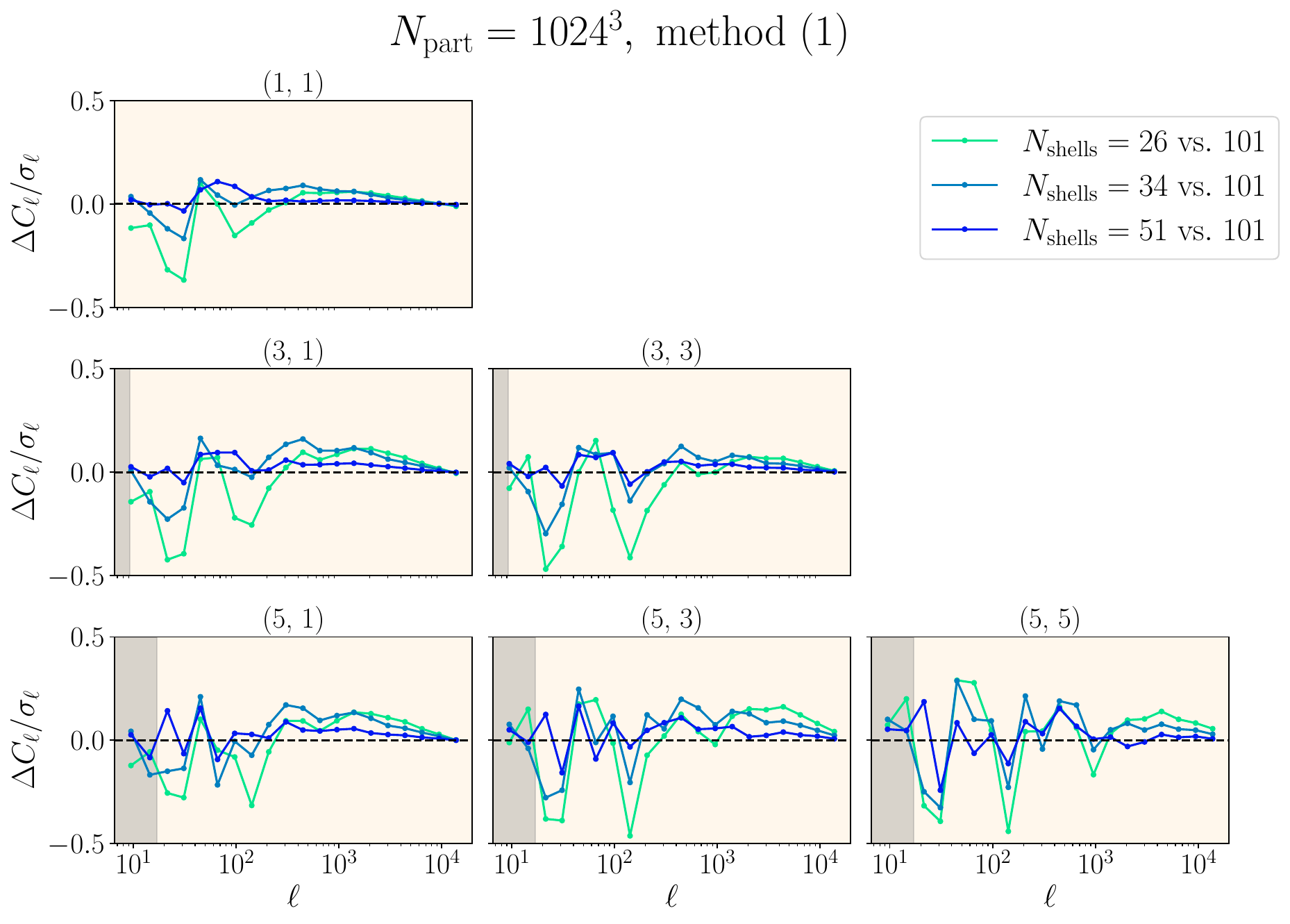}
    \caption{Convergence of the \methodone algorithm (method \ref{item:method1}) with the number of shells. The plot shows $\Delta C_\ell\equiv C_\ell^{\nshells}-C_\ell^{101}$ of the tomographic $\kappa$ maps for $\nshells = 26$, 34 and 51, normalized by the diagonal of the Gaussian covariance matrix (for the first, third and fifth redshift bins). Gray-shaded areas indicate scales larger than the size of the box. Analogous to \cref{fig:Cell_all_method1} but for $\npart = 1024^3$.}
    \label{fig:Cell_all_method1_1024}
\end{figure*}

\begin{figure*}
    \centering
    \includegraphics[width=0.95\textwidth]{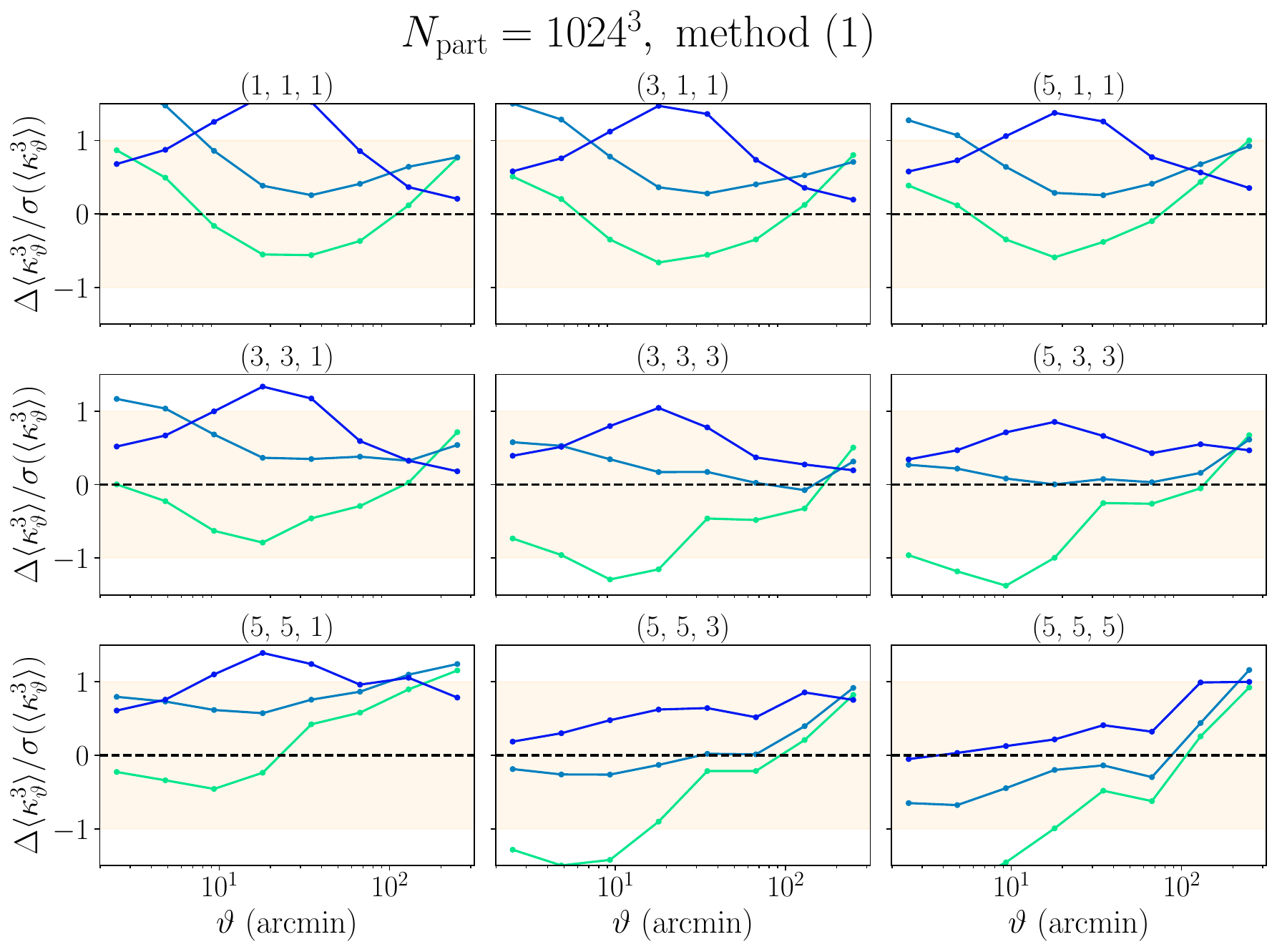}
    \includegraphics[width=0.95\textwidth]{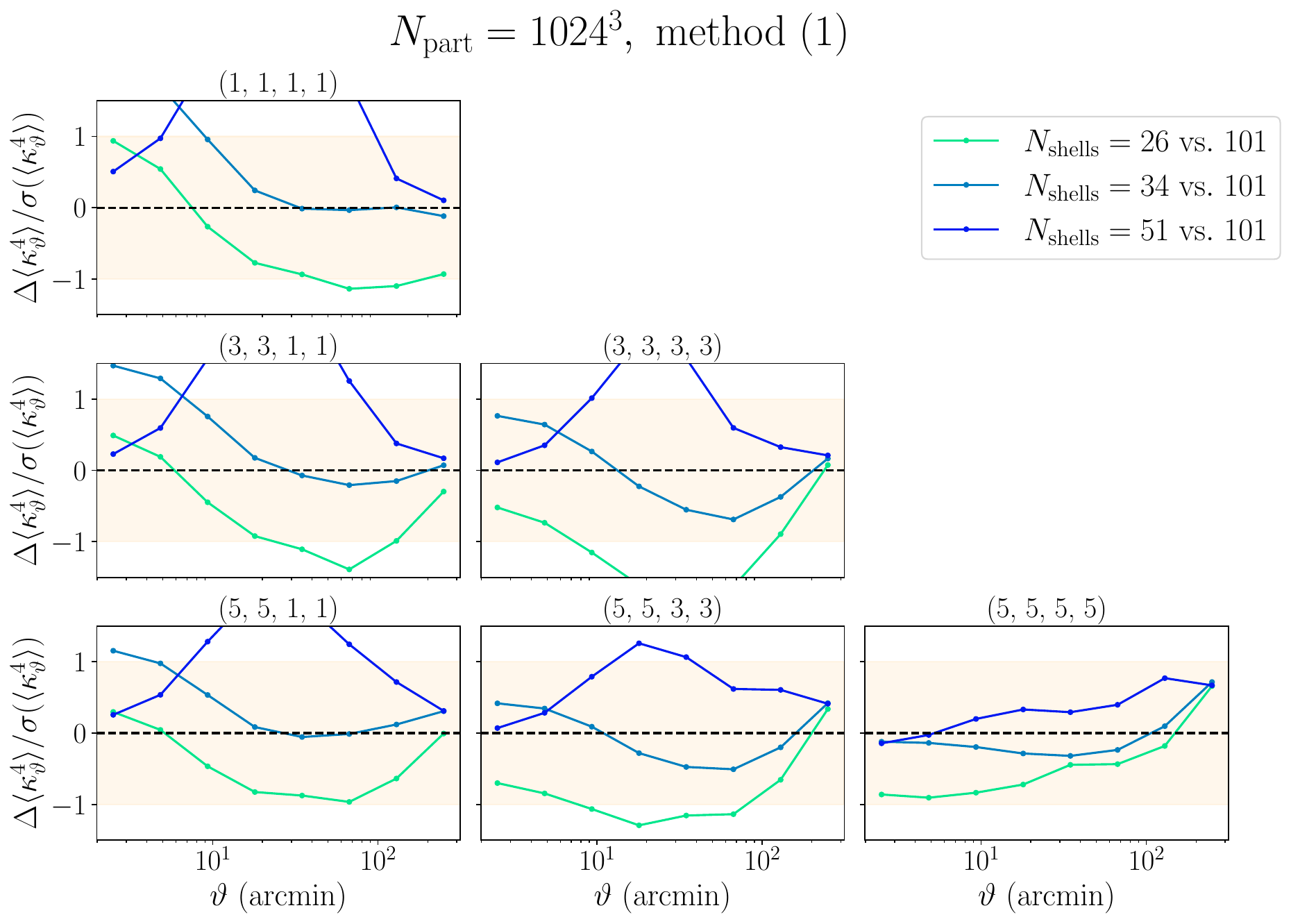}
    \caption{Convergence of the \methodone algorithm (method \ref{item:method1}) with the number of shells. The plot shows $\Delta \langle\kappa^n_\vartheta\rangle\equiv \langle\kappa^n_\vartheta\rangle^{\nshells}-\langle\kappa^n_\vartheta\rangle^{101}$ for $\nshells = 26$, 34 and 51 ($n=3$ on top, $n=4$ on bottom), normalized by the standard deviation of the ten $\nshells = 101$ realizations (for the first, third and fifth redshift bins). The orange-shaded areas indicate the $1\sigma$ regions. Analogous to \cref{fig:kappa_all_method1} but for $\npart = 1024^3$.}
    \label{fig:kappa_all_method1_1024}
\end{figure*}

In this appendix, we show the convergence of our $\npart = 1024^3$ simulations as a function of \nshells. We do this for the same weak lensing statistics as in the main body of the paper (\cref{sec:results_method1_1024}):
\begin{itemize}
    \item Angular power spectrum. In \cref{fig:Cell_all_method1_1024}, we show the $\Delta C_\ell\equiv C_\ell^{\nshells}-C_\ell^{101}$ of the tomographic $\kappa$ maps for $\nshells = 26$, 34 and 51, normalized by the diagonal of the Gaussian covariance matrix. We find that the $C_\ell$ converge to the $\nshells = 101$ case as \nshells increases, just like in the $\npart = 2048^3$ case, see \cref{fig:Cell_all_method1}.
    \item Higher-order moments of $\kappa$. In \cref{fig:kappa_all_method1_1024}, we show the same convergence test as in \cref{fig:Cell_all_method1_1024} but for the third (top) and fourth (bottom) moments of $\kappa$. Unlike the $C_\ell$ case, where Gaussian covariance matrices were used in the denominator, the measurements shown here are normalized by the standard deviation estimated from the ten realizations (calculated for $\nshells = 101$). We find that, unlike in the $\npart = 2048^3$ case shown in \cref{fig:kappa_all_method1}, we do not see a clear trend towards convergence as \nshells increases.
\end{itemize}

\acknowledgments

This paper underwent internal review within the LSST Dark Energy Science Collaboration. We thank the internal reviewers, Judit Prat and Natalia Porqueres, for their constructive comments, which helped shape and improve the manuscript.

Author contributions: 
JMF contributed to the development of \pollux, produced the lightcone simulations, measured the two-point and higher-order statistics on the simulations and produced plots and tables.
CD supervised the development of this project, developed \pollux, studied the optimal binning of the lightcones and contributed to the writing of the manuscript.
JHD helped with designing the $N$-body simulations and the ray-tracing calculations.
CC supervised the development of this project and provided useful feedback on the analysis and the writing of the manuscript.
KH, PL and NF contributed to the production of the $N$-body simulations and provided useful feedback on the writing of the manuscript.
AB provided useful discussions during the development of this analysis.
SSN, LC and CU contributed to the $\kappa$-PDF analysis and to the writing of the corresponding appendix.

The DESC acknowledges ongoing support from the Institut National de Physique Nucl\'eaire et de Physique des Particules in France; the Science \& Technology Facilities Council in the United Kingdom; and the Department of Energy and the LSST Discovery Alliancein the United States.  DESC uses resources of the IN2P3 Computing Center (CC-IN2P3--Lyon/Villeurbanne - France) funded by the Centre National de la Recherche Scientifique; the National Energy Research Scientific Computing Center, a DOE Office of Science User Facility supported by the Office of Science of the U.S.\ Department of Energy under Contract No.\ DE-AC02-05CH11231; STFC DiRAC HPC Facilities, funded by UK BEIS National E-infrastructure capital grants; and the UK particle physics grid, supported by the GridPP Collaboration. This work was performed in part under DOE Contract DE-AC02-76SF00515.

Work at Argonne National Laboratory was supported by the U.S. Department of Energy, Office of High Energy Physics. This research used resources of the Argonne Leadership Computing Facility, a U.S. Department of Energy (DOE) Office of Science user facility at Argonne National Laboratory and is based on research supported by the U.S. DOE Office of Science-Advanced Scientific Computing Research Program. Argonne, a U.S. Department of Energy Office of Science Laboratory, is operated by UChicago Argonne LLC under contract no. DE-AC02-06CH11357. LC and CU were supported by the European Union (ERC StG, LSS\_BeyondAverage, 101075919). SSN acknowledges financial support by SECIHTI grant CBF 2023-2024-162 and from DGAPA-PAPIIT-UNAM grant No. IG102123, as well as funding for long-term international research stays provided by UNAM.


\bibliographystyle{JHEP}
\bibliography{biblio.bib}

\end{document}

%% file: DESC-PUB-00233_author_list.tex
\author[1,2]{{J. Mena-Fern\'andez},}
\author[1]{{C.~Doux},}
\author[3]{{J.~Harnois-D\'eraps},}
\author[4]{{K.~Heitmann},}
\author[1]{{C.~Combet},}
\author[4]{{P.~Larsen},}
\author[4]{{N.~Frontiere},}
\author[5]{{A.~Bera},}
\author[6]{{S.~Samario-Nava},}
\author[7]{{L.~Castiblanco},}
\author[7,3]{{C.~Uhlemann},}
\collaboration{{The LSST Dark Energy Science Collaboration}}

\affiliation[1]{Univ. Grenoble Alpes, CNRS, LPSC-IN2P3, 38000 Grenoble, France}
\affiliation[2]{Centre de Physique des Particules de Marseille, Aix Marseille Univ, CNRS/IN2P3, CPPM, Marseille, France}
\affiliation[3]{School of Mathematics, Statistics and Physics, Newcastle University, Herschel building, NE1 7RU, Newcastle-upon-Tyne, UK}
\affiliation[4]{High Energy Physics Division, Argonne National Laboratory, 9700 S Cass Ave, Lemont, IL 60439, USA}
\affiliation[5]{Department of Physics, The University of Texas at Dallas, Richardson, TX 75080, USA}
\affiliation[6]{Universidad Nacional Aut\'onoma de M\'exico, Cuernavaca Morelos 62210, Mexico}
\affiliation[7]{Universit\"at Bielefeld, Postfach 100131, 33501 Bielefeld, Germany}